\documentclass[preprint,showpacs,preprintnumbers,amsmath,amssymb]{revtex4}
\usepackage{graphicx}
\usepackage{dcolumn}
\usepackage{bm}
\usepackage{array,tabularx}      
\usepackage{amsmath}
\usepackage{mathrsfs} 
\usepackage{indentfirst}
\usepackage{slashed}

\begin{document}
\title{The two-pseudoscalar-meson decay of $\chi_{cJ}$ with twist-3 corrections}
\author{Ming-Zhen Zhou}%
 \email{zhoumz@mail.ihep.ac.cn}
\affiliation{School of Physical Science and Technology, Southwest
University, Chongqing 400715,~~People's Republic of
China\footnote{Mailing address.} }
\author{Hai-Qing Zhou}
 \email{zhouhq@mail.ihep.ac.cn}
  \affiliation{Department of Physics, Southeast University,Nanjing 211189 China}

\date{\today}
\begin{abstract}
The decays of $\chi_{cJ} \rightarrow \pi^+\pi^-, K^+K^-$ $(J=0,2)$
are discussed within the standard and modified hard scattering
approach when including the contributions from twist-3 distribution
amplitudes and wave functions of the light pseudoscalar meson. A
model for twist-2 and twist-3 distribution amplitudes and wave
functions of the pion and kaon with BHL prescription are proposed as
the solution to the end-point singularities. The results show that
the contributions from twist-3 parts are actually not power
suppressed comparing with the leading-twist contribution. After
including the effects from the transverse momentum of light meson
valence-quark state and Sudakov factors, the decay widths of the
$\chi_{cJ}$ into pions or kaons are comparable with the their
experimental data.
\end{abstract}

\maketitle

\section{Introduction}
The factorization form in the framework of the hard-scattering
picture \cite{HSA} is usually used in hadronic processes with large
momentum transfer. In this picture, the full amplitude is factorized
as a convolution of process independent distribution amplitudes of
hadrons and process-dependent hard-scattering amplitude. However,
the applicability of this approach at experimentally accessible
momentum transfers, typically a few GeV, is
questionable\cite{QustEP}. One of the reasons is the large
contributions from the soft end-point regions. A modified
perturbative approach, so-call modified hard-scattering
approach(mHSA), has been proposed by Li and Sterman
\cite{LiSterman}, where quark transverse momenta and Sudakov
suppressions are taken into account. The advantage of this modified
perturbative approach is the strong suppression of the soft
end-point regions where the pQCD can not be applied.

The exclusive charmonium decays have attracted interest for decades
as they are an excellent laboratory for studying quark-gluon
dynamics at relatively low energies. In the decay of P-wave
charmonium $\chi_{c0,2}$ to a pair of pseudoscalars, one finds that
the lowest Fock state, the color-singlet contribution, alone is not
sufficient to accommodate the experimental data. This discrepancy
provides an important arena in which to test our understanding of
the boundary domain between perturbative and non-perturbative QCD.
From a naive point of view, QCD radiative correction is suppressed
by the factor $\alpha_s(Q^2)$ and the contribution from higher Fock
states, higher twist distribution amplitudes are suppressed by the
factor $1/Q^2$. But this is not always the real case. In Ref.
\cite{octet}, the author showed that in the decay of $\chi_{c0,2}
\rightarrow PP$ (where P represents a light pseudoscalar meson.) the
color-octet contribution from the higher Fock state contributes the
same level as the color singlet state.

To reveal the decay mechanism more clear, the systemic reanalysis on
the contributions from higher twist distribution amplitudes will be
significative and interested. When the amplitude of a physical
process with large momentum transfer $Q^2$ is only related to one
hadron wave function, there is a suppression for the contribution
from higher Fock states and higher twist distribution amplitudes.
For example, the dominating contribution of $\pi-\gamma$ transition
form factor \cite{tranF} comes from the leading twist distribution
amplitude of valence quark state. The QCD correction is only about
$10\%-20\%$ \cite{alphastranF} and corrections from higher Fock
state and higher twist distribution amplitudes \cite{hightw} are
suppressed by additional powers of ¦Á$1/Q^2$. However, for the
exclusive processes with the overlap of the wave functions of the
initial hadron and final hadrons there are a lot of space to discuss
the contributions from higher twist terms. For instance, the
contribution of twist-3 distribution amplitudes to the pion
electromagnetic form factor is comparable and even larger than the
contribution from the leading-twist distribution amplitude of the
pion at intermediate energy region of $Q^2$, being $2-40~GeV^2$ in
Ref. \cite{pionF}. The similar result has also been obtained for the
kaon electromagnetic form factor in Ref. \cite{KaonF}. More
discussion about the contributions from the higher twist
distribution amplitudes can be found in the Heavy-to-light
transition form factors \cite{HLF}, the nonleptonic two-body decays
of the B meson \cite{BPP}, and so on.

In this paper, we apply both the mHSA and the sHSA to reexamine the
decay of $\chi_{cJ} \rightarrow \pi^+\pi^-, K^+K^-$ $(J=0,2)$
including the contribution from the two-particle twist-3 wave
functions and distribution amplitudes of the light pseudoscalar
meson. In the sHSA case, the end-point singularity can be avoided by
using the  BHL prescription \cite{BHL} in which the light meson
distribution amplitudes are rewritten with exponential suppression
factors. Our results show that the contributions from twist-3 wave
function  are comparable with or even larger than the leading twist
contribution. Comparing with \cite{octet}, the present analysis has
an advantage that the theoretical uncertainty from light meson
higher twist wave functions is less than that from the color octet
wave function and the constituent gluon from the octet state of the
charmonium. One can find the discussion on light meson twist-3 wave
functions in many literatures \cite{DA1,DA2,DA3}.

The paper is organized as following. In section II the main
calculation of the hard-scattering amplitude in the modified
perturbative QCD approach is presented. In section III, we present
our model for the light meson two-particle twist-2 and twist-3 wave
functions and the distribution amplitude within the BHL scheme. The
section IV is the  numerical analysis and section IV is our
conclusion. The coefficients of hard-scattering amplitudes in the
momentum space and in the b space are given in Appendix A and B. The
Sudakov factor is presented in Appendix C.

\section{Calculation of hard-scattering amplitudes}

\begin{figure}[ht]
\begin{center}
\setlength{\unitlength}{1cm}
\begin{minipage}[t]{7.5cm}
\begin{picture}(7.0,5.5)
\includegraphics*[scale=0.5,angle=0.]{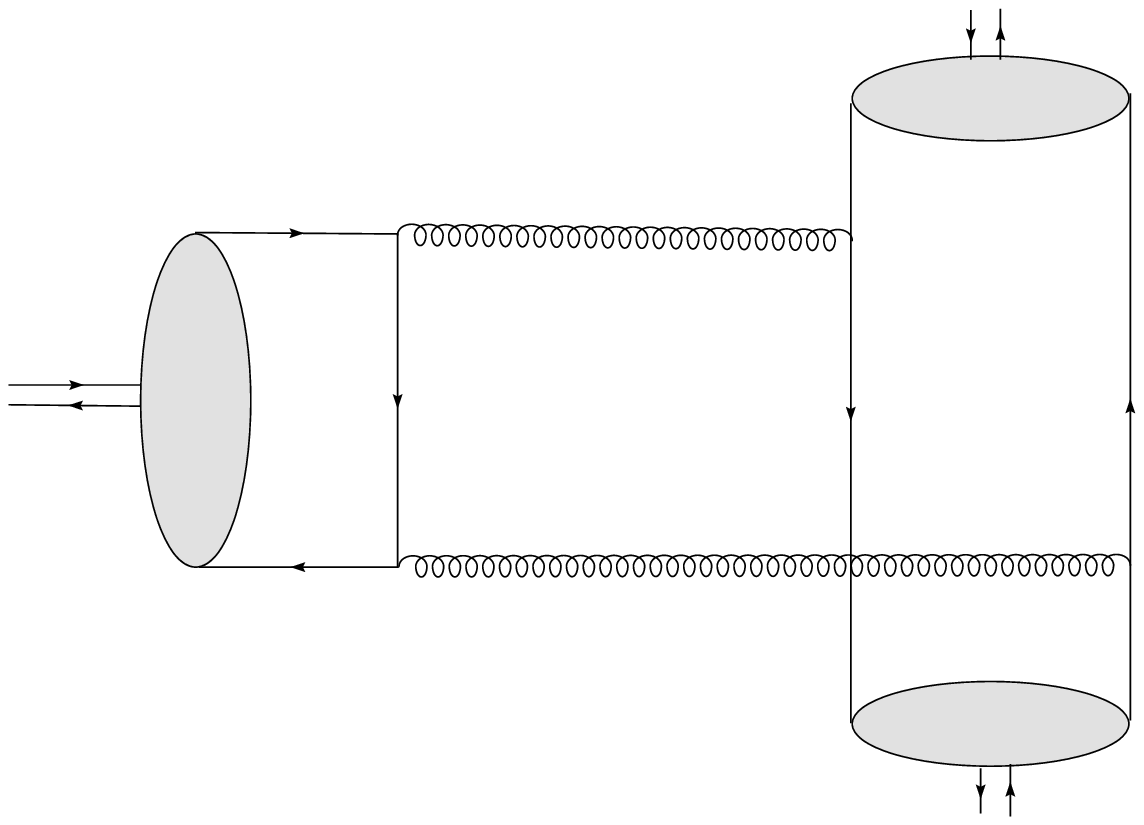}
\end{picture}\par
\begin{center}
(a)
\end{center}
\end{minipage}
\hfill 
\begin{minipage}[t]{7.5cm}
\begin{picture}(7.0,5.5)
\includegraphics*[scale=0.5,angle=0.]{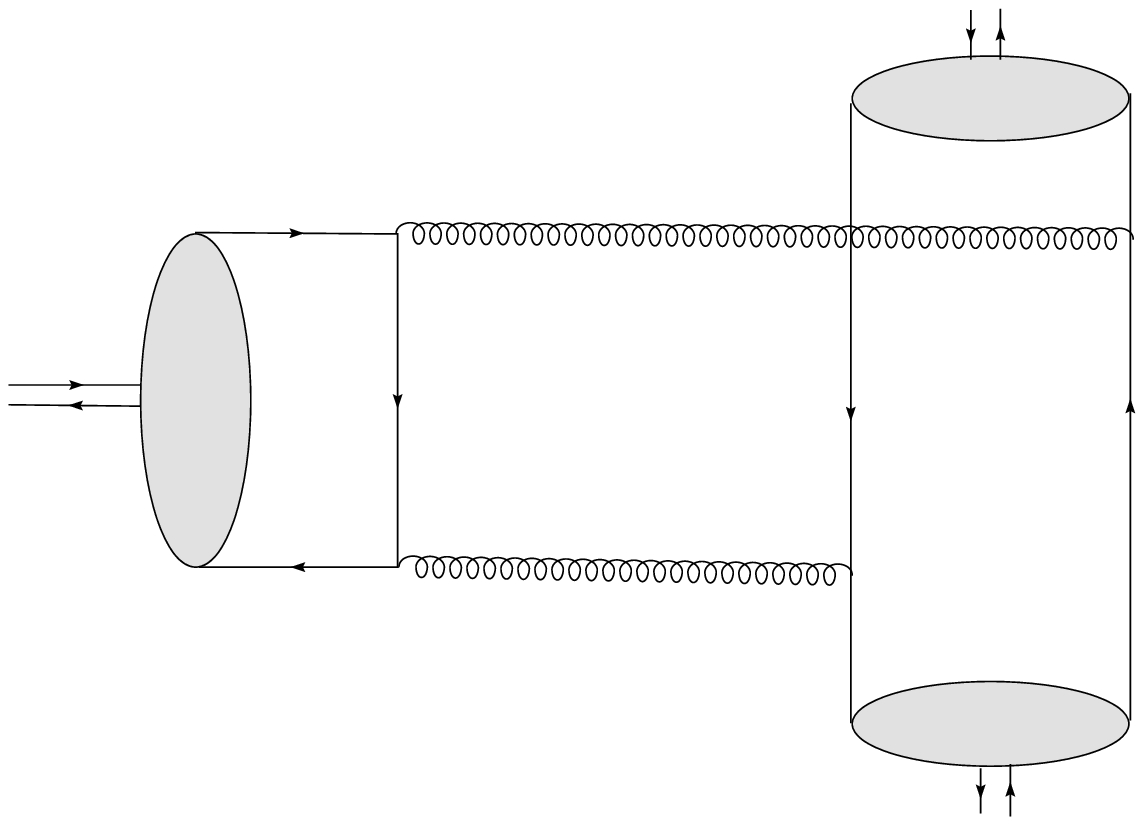}
\end{picture}\par
\begin{center}
(b)
\end{center}
\end{minipage}
\caption{The hard scattering diagrams for the decay
$\chi_{cJ}\rightarrow \pi^+\pi^-, K^+K^-$(J=0,2).}
\end{center}
\end{figure}

The two-meson decays of $\chi_{cJ}$ can be described as $c\bar{c}$
quarks annihilate into two gluons and then materializing into two
final-state mesons as illustrated in Fig.1. We work in the rest
frame of $\chi_{cJ}$ meson and take approximation $M=2 m_c $ where M
is the mass of the charmonium meson and $m_c$ is the mass of
c-quark. The masses of pion and kaon are canceled in the Chiral
limit. Under these conventions, the momentums of the initial- and
final-state mesons are described in terms of light-cone variables as
\begin{eqnarray}
P=({M\over \sqrt{2}},{M\over \sqrt{2}},0_T),~~~p_1=({M\over
\sqrt{2}},0,0_T),~~~p_2=(0,{M\over \sqrt{2}},0_T),
\end{eqnarray}
where P is the momentum of the charmonium and $p_1$ and $p_2$ are
the momentums of $\pi^+(K^+)$ and $\pi^-(K^-)$, respectively. In the
following, we only present the two-pion decay of $\chi_{cJ}$ for the
two-kaon decay is similar.

For the two-pion decay of $\chi_{cJ}$, its decay width can be
written as:
\begin{equation}
\Gamma[\chi_{cJ}\rightarrow \pi^+\pi^-]={1 \over 16 \pi M}{1\over
2J+1}\sum_{J_z} |\mathcal {M}(\chi_{cJ}\rightarrow \pi^+\pi^-)|^2,
\end{equation}
where $\mathcal {M}$ is the transitional matrix element. As usual,
this matrix element within the mHSA can be factorized as the
convolution with respect to the momentum fractions $x, y$ and
transverse separation scales $\textbf{b}_1$, $\textbf{b}_2$ of the
two pions,
\begin{eqnarray}
\mathcal {M}(\chi_{cJ}\rightarrow \pi^+\pi^-)&=&\int_0^1 dx dy \int
{d^2\textbf{b}_1 \over 4\pi}{d^2\textbf{b}_2 \over 4\pi} \nonumber \\
&&\times\sum_{i,j}\Psi_i(y,\textbf{b}_2)\mathcal
{T}_{HJ}^{ij}(x,y,\textbf{b}_1,
\textbf{b}_2)\Psi_j(x,\textbf{b}_1)e^{-S(x,y,\textbf{b}_1,\textbf{b}_2,t_1,t_2)},
\end{eqnarray}
here the scripts ($i,j=\pi,p,\sigma$) mean the twist-2 and twist-3
wave functions for the pion meson and S is the Sudakov factor.
$\mathcal {T}_{HJ}^{ij}(x,y,\textbf{b}_1, \textbf{b}_2)$ is the
Fourier transform of the hard-scattering amplitudes
$T_{HJ}^{ij}(x,y,\textbf{k}_1, \textbf{k}_2)$
\begin{eqnarray}
\mathcal {T}_{HJ}^{ij}(x,y,\textbf{b}_1,
\textbf{b}_2)&=&\int{d^2\textbf{k}_1 \over (2\pi)^2}{d^2\textbf{k}_2
\over (2\pi)^2}T_{HJ}^{ij}(x,y,\textbf{k}_1, \textbf{k}_2)e^{-i
\textbf{k}_1\cdot\textbf{b}_1-i \textbf{k}_2\cdot\textbf{b}_2}.
\end{eqnarray}
$T_{HJ}^{ij}(x,y,\textbf{k}_1, \textbf{k}_2)$ can be calculated from
the graphs shown in Fig.1 and the scripts of nonzero terms are
$ij=\pi\pi,pp,p\sigma,\sigma p,\sigma\sigma$, $\textbf{k}_1$ and
$\textbf{k}_2$ are the intrinsic transverse momentum of two
final-state pion mesons, respectively.

To calculate the transitional matrix element, the wave functions of
the pion meson and the charmonium should be introduced. Similar to
the definition of distribution amplitudes with leading and
next-to-leading twist \cite{DA1}, the light-cone wave functions of
pion are defined in terms of bilocal operator matrix element
\begin{eqnarray}
&&<\pi(p_\pi)|\bar{q}_\beta(z_1)q_\alpha(z_2)|0>={i f_\pi \over
4}\int_0^1 dx \int {d^2\textbf{k}_\perp\over 16 \pi^3}~ e^{i(x
p_\pi\cdot z_1+(1-x)p_\pi\cdot
z_2-\textbf{k}_\perp\cdot(\textbf{z}_1-\textbf{z}_2)_\perp)}\nonumber
\\
&&~~~~~~~~~~~~~~~~~~~ \times \{\slashed{p}_\pi \gamma_5
\Psi_\pi(x,\textbf{k}_\perp)-\mu_\pi
\gamma_5(\Psi_p(x,\textbf{k}_\perp)-\sigma_{\mu\nu}p_\pi^\mu
(z_1-z_2)^\nu {\Psi_\sigma(x,\textbf{k}_\perp)\over 6} )
\}_{\alpha\beta},
\end{eqnarray}
where $f_\pi$ is the decay constant of pion and the parameter
$\mu_\pi = m_\pi^2/(m_u + m_d)$ for charged pion. $\Psi_\pi$,
$\Psi_p$ and $\Psi_\sigma$ are the twist-2 and twist-3 wave
functions, respectively. The twist-3 wave functions contributes
power corrections. But at $m_c$ energy scale, the Chirally enhanced
parameter $r_\pi = \mu_\pi/m_c \sim 1$ is large enough to consider
the twist-3 contributions in charmonium decays.

For simplif, the wave function of the charmonium is taken as
\cite{BSform},
\begin{eqnarray}
\chi(P,q;J,J_z)=\sum_{M,S_z}2\pi \delta(q^0-{\vec{q}^2 \over 2
m_c})<LM;SS_z|JJ_z>\psi_{LM}(\vec{q})\mathcal {P}_{SS_z}(P,q),
\end{eqnarray}
where q is the relative momentum between the quark and anti-quark,
$\psi_{LM}(\vec{q})$ and $\mathcal {P}_{SS_z}(P,q)$ are the spatial
wave function and spin projection operators, respectively. The spin
projection operators $\mathcal {P}_{SS_z}(P,q)$ are
\begin{eqnarray}
\mathcal {P}_{SS_z}(P,q)=\sqrt{{3 \over 8 m_c^3}}[
m_c+({\slashed{P}\over 2}+\slashed{q})](1+{\slashed{P}\over
M})\Pi_{SS_z}[m_c-({\slashed{P}\over 2}-\slashed{q})]
\end{eqnarray}
with $\Pi_{SS_z}=-\gamma_5$ for $S=0$ and
$\Pi_{SS_z}=-\slashed{\varepsilon}(S_z)$ for $S=1$. Here
$\varepsilon(S_z)$ refers to the spin part of the wave function,
$\varepsilon(S_z)=(0,0,0,1)$ for $S_z=0$ and
$\varepsilon(S_z)=(0,\mp1,-i,0)/\sqrt{2}$ for $S_z=\pm1$.

With the definition of the wave functions for the initial- and
final-state hadrons, the transition matrix element of the two-pion
decay for $\chi_{cJ}$ can be calculated in the coordinate space by
standard method and then hard-scattering amplitudes
$T_{HJ}^{ij}(x,y,\textbf{k}_1,\textbf{k}_2)$ can be extracted as
\begin{eqnarray}
T_{HJ}^{ij}(x,y,\textbf{k}_1,\textbf{k}_2)=-{32 \pi^2 \over 27}
\alpha_s(t_1^2)\alpha_s(t_2^2)\hat{T}_{\mu\nu}^{ij} \int {d^4 q\over
(2 \pi)^4} {Tr[\chi(P,q;J,J_z) \hat{\mathcal {O}}^{\mu\nu}(q)]\over
(l_1^2+i \epsilon)(l_2^2+i \epsilon)}+(l_1 \leftrightarrow l_2),
\end{eqnarray}
where $l_1=x p_1+y p_2+\textbf{k}_1-\textbf{k}_2$ and
$l_2=(1-x)p_1+(1-y)p_2-\textbf{k}_1+\textbf{k}_2$ are the momentum
of two intermediate gluons, respectively. The operator
$\hat{\mathcal {O}}^{\mu\nu}(q)$, which is related to the two-gluon
annihilation of charmonium, reads
\begin{eqnarray}
\hat{\mathcal
{O}}^{\mu\nu}(q)={\gamma^\mu(\slashed{k}_c+m_c)\gamma^\nu \over
k_c^2-m_c^2+i \epsilon},
\end{eqnarray}
where $k_c=(l_1-l_2)/2+q$ is the corresponding momentum of the
c-quark propagator. The operators $\hat{T}_{\mu\nu}^{ij}$ are
related to the two-meson materialization of two gluons with
different twist wave functions and are expressed as
\begin{eqnarray}
&&\hat{T}_{\mu\nu}^{\pi\pi}={f_\pi^2 \over 16}Tr[\gamma_\mu
\slashed{p}_2\gamma_5\gamma_\nu\slashed{p}_1\gamma_5],~~\hat{T}_{\mu\nu}^{pp}={f_\pi^2
\mu_\pi^2 \over 16}Tr[\gamma_\mu \gamma_5\gamma_\nu\gamma_5],
\nonumber \\
&& \hat{T}_{\mu\nu}^{p\sigma}=-i{f_\pi^2 \mu_\pi^2 \over 4\times
24}Tr[\gamma_\mu \gamma_5\gamma_\nu
\sigma_{\alpha\beta}\gamma_5]p_1^\alpha({\partial \over
\partial l_{1\beta}}-{\partial \over
\partial l_{2\beta}}),  \nonumber \\
&& \hat{T}_{\mu\nu}^{\sigma p}=i{f_\pi^2 \mu_\pi^2 \over 4\times
24}Tr[\gamma_\mu \sigma_{\alpha\beta}\gamma_5\gamma_\nu
\gamma_5]p_2^\alpha({\partial \over
\partial l_{1\beta}}-{\partial \over
\partial l_{2\beta}}),  \nonumber \\
&& \hat{T}_{\mu\nu}^{\sigma \sigma}=-{f_\pi^2 \mu_\pi^2 \over
16\times 36}Tr[\gamma_\mu
\sigma_{\tau\alpha}p_2^\tau\gamma_5\gamma_\nu
\sigma_{\delta\beta}p_1^\delta\gamma_5]({\partial \over
\partial l_{1\alpha}}-{\partial \over
\partial l_{2\alpha}})({\partial \over
\partial l_{1\beta}}-{\partial \over
\partial l_{2\beta}}),
\end{eqnarray}
where the partial $({\partial \over
\partial l_{1\nu}}-{\partial \over
\partial l_{2\nu}})$ comes from the $(z_1-z_2)^\nu$ term of Eq.(5).
Here we don't take the momentum projection of Eq.(5), which can be
obtained by transforming the parameters in terms of coordinate
variable in Eq.(5) into the momentum space configuration, in
Ref.\cite{Li} or \cite{Beneke}.

For P-wave charmonium decay, the dominant contribution is given by
$q^\lambda$ term of $Tr[\chi(P,q;J,J_z) \hat{\mathcal
{O}}^{\mu\nu}(q)]$ and the result after the integration of momentum
$q$ can be rewritten as
\begin{eqnarray}
\sum_{M,S_z} \sqrt{{3\over \pi}}{R_p^{'}(0)\over 4}\left \langle
1M,1S_z|JJ_z\right \rangle \varepsilon^\lambda(M)
Tr\left[\hat{\mathcal {O}}^{\mu\nu}_\lambda(0)
\mathscr{P}_{1S_z}(P,0)+\hat{\mathcal
{O}}^{\mu\nu}(0)\mathscr{P}_{1S_z,\lambda}(P,0)\right],
\end{eqnarray}
with the definition of the partial of the P-wave function at the
origin
\begin{eqnarray}
\int {d^3q \over (2 \pi)^3} q^\lambda
\psi_{1M}(\vec{q})=\sqrt{{3\over \pi}} {R_p^{'}(0)\over
4}\varepsilon^\lambda(M),
\end{eqnarray}
where $\varepsilon^\lambda(M)$ refers to the orbital part of the
wave function and
\begin{eqnarray}
&&~~~~\hat{\mathcal {O}}^{\mu\nu}(0)={\gamma^\mu[
(\slashed{l}_1-\slashed{l}_2)/2+m_c]\gamma^\nu \over
(l_1-l_2)^2/4-m_c^2},
\nonumber \\
&&~~~~\hat{\mathcal {O}}^{\mu\nu}_\lambda(0)={\gamma^\mu
\gamma_\lambda \gamma^\nu \over
(l_1-l_2)^2/4-m_c^2}-{(l_1-l_2)_\lambda \gamma^\mu [(\slashed{l}_1-\slashed{l}_2)/2+m_c]\gamma^\nu \over [(l_1-l_2)^2/4-m_c^2]^2}, \nonumber \\
&&~~~~\mathcal {P}_{1S_z}(P,0)=\sqrt{{3 \over 8 m_c^3}}(
m_c+{\slashed{P}\over 2})(1+{\slashed{P}\over
M})\Pi_{1S_z}(m_c-{\slashed{P}\over 2}),
\nonumber \\
&&~~~~\mathcal {P}_{1S_z,\lambda}(P,0)=\sqrt{{3 \over 8
m_c^3}}[\gamma_\lambda (1+{\slashed{P}\over
M})\Pi_{1S_z}(m_c-{\slashed{P}\over 2})+ ( m_c+{\slashed{P}\over
2})(1+{\slashed{P}\over M})\Pi_{1S_z}\gamma_\lambda ].
\end{eqnarray}

The final hard-scattering amplitudes with transverse momentum can be
expressed as
\begin{equation}
T_{HJ}^{I}(x,y,\textbf{k}_1,\textbf{k}_2)={16 \over 9
\sqrt{6}}C^I({\mu_\pi^2\over m_c^2})^{t_I-2} N_0 \sigma_J
{\alpha_s(t_1^2)\alpha_s(t_2^2) \over D_1^{d_I} D_2^{d_I}
N^{n_I}}\sum_{i=0}^{i_I}C_i^I(J)N^i,
\end{equation}
where $N_0=16 \pi^{3/2}f_\pi^2 m_c^{5/2}R_p^{'}(0)$, $\sigma_0=1$
for $\chi_{c0}$ and $\sigma_2=1/\sqrt{2}$ for $\chi_{c2}$. The
coefficients $C^I,~t_I,~d_I,~n_I,~i_I$ are listed in Table.I. The
virtualities of the internal c-quark and the two intermediate gluons
are
\begin{eqnarray}
&&D_1=4 m_c^2 x y-\textbf{K}^2+i\epsilon, \nonumber \\
&&D_2=4 m_c^2 (1-x)(1-y)-\textbf{K}^2+i\epsilon \nonumber \\
&&N=2 m_c^2 (x+y-2xy)+\textbf{K}^2-i\epsilon
\end{eqnarray}
with $\textbf{K}=\textbf{k}_1-\textbf{k}_2$. The coefficients
$C_0^{\pi\pi}(J)=1$ and $C_1^{\pi\pi}(J)=(-2)^{J/2} m_c^2 (x - y)^2$
are in agreement with the results from Ref.\cite{octet} and the
others coefficients are showed in Appendix A.
\begin{table}
\caption{The values of coefficients $C^I,~t_I,~d_I,~n_I,~i_I$ in the
eq.(14).}
\begin{ruledtabular}
\begin{tabular}{cccccc}
 I & $\pi\pi$ & $pp$ & $p\sigma $
 & $\sigma p$ & $\sigma\sigma$ \\
\hline $C^I$ & 1 & ${1\over 4}$&${1\over 3}$&${1\over 3}$&${1\over
9}$ \\
$t_I$& 2&3&3&3&3 \\
$d_I$&1&1&2&2&3 \\
$n_I$&2&2&3&3&4 \\
$i_I$&1&1&3&3&6 \\
\end{tabular}
\end{ruledtabular}
\end{table}

Next, the fact that hard-scattering amplitudes $T_{HJ}^{ij}$ depend
on $\textbf{k}_1$ and $\textbf{k}_2$ only in the combination
$\textbf{K}$ implies the following result for the Fourier transform
of them
\begin{eqnarray}
\mathcal {T}_{HJ}^{I}(x,y,\textbf{b}_1,\textbf{b}_2)&=&{2 \over 9
\sqrt{6}}C^I({\mu_\pi^2\over m_c^2})^{t_I-2} \mathcal {N}_0 \sigma_J
\delta(\textbf{b}_1-\textbf{b}_2)
\alpha_s(t_1^2)\alpha_s(t_2^2)\nonumber \\
&&[i(\sum_{i=0}^{d_I-1}A_{1i}^I(J)H_i^{(1)}(r_1)-\sum_{i=0}^{d_I-1}A_{2i}^I(J)H_i^{(1)}(r_2))+\sum_{i=0}^{n_I-1}B_i^I(J)K_i(r_3)],
\end{eqnarray}
where $\mathcal {N}_0=16 \pi^{3/2}f_\pi^2 R_p^{'}(0)/m_c^{3/2}$,
$r_1=\sqrt{x y}b$, $r_2=\sqrt{(1-x)(1-y)}b$ and $r_3=\sqrt{(x+y-2 x
y)/2}b$ with $b=2 m_c b_1$. $H_i^{(1)}$ and $K_i$ denote Hankel and
modified Bessel functions, respectively. The $\delta$-function,
which simplifies the numerical work enormously, means that the two
pions emerge from the decay with identical transverse separations.
The coefficients $A_{1i}^I(J),~A_{2i}^I(J)$ and $B_i^I(J)$ are
listed in Appendix B.

The novel ingredient of the mHSA is the Sudakov factor $e^{-S}$,
which takes into account those gluonic radiative corrections not
accounted for in the QCD evolution of the wave function. In
next-to-leading-log approximation, the Sudakov exponent reads
\begin{eqnarray}
S(x,y,\textbf{b}_1,\textbf{b}_2,t_1,t_2)&=&s(x,b_1,2
m_c)+s(1-x,b_1,2 m_c)+s(y,b_2,2 m_c)+s(1-y,b_2,2 m_c)\nonumber \\
&&-{1 \over \beta_1}\ln
{\ln(t_1/\Lambda_{QCD})\ln(t_2/\Lambda_{QCD})\over
\ln(1/(b_1\Lambda_{QCD}))\ln(1/(b_2\Lambda_{QCD}))},
\end{eqnarray}
where the function $s(x,b_i,Q)$ with next-to-leading-log correction
is given in Appendix C. The last term in Eq.(17) arises from a
renormalization group transformation from the factorization scales
$\mu_{Fi}$ to the renormalization scales $t_j$ at which the
hard-scattering amplitudes $\mathcal
{T}_{HJ}^{I}(x,y,\textbf{b}_1,\textbf{b}_2)$ are evaluated.

The renormalization scales appearing in $\alpha_s$ and in the
Sudakov exponent are chosen as
\begin{eqnarray}
&&t_1=max\{2 m_c \sqrt{x y},1/b_1,1/b_2\}, \nonumber \\
&&t_2=max\{2 m_c \sqrt{(1-x)(1-y)},1/b_1,1/b_2\}
\end{eqnarray}
by the virtualities of the intermediate gluons, which depend
non-trivially on the integration variables. This choice of the
renormalization scale avoids large logs from higher-order pQCD. The
factorization scale is given by the quark-antiquark separation
$\textbf{b}_i$, $\mu_{Fi}=1/b_i$. The ratio $1/b_i$ marks the
interface between non-perturbatively soft momenta, which are
implicitly accounted for in the pion wave functions, and the
contributions from semi-hard gluons, incorporated in a perturbative
way in the Sudakov factor.

Replacing $e^{-S}$ by 1 and ignoring the transverse momenta in
$T_{HJ}^{I}(x,y,\textbf{k}_1,\textbf{k}_2)$, one finds the decay
amplitude within the sHSA as derived by Duncan and Mueller
\cite{DM},
\begin{eqnarray}
\mathcal {M}(\chi_{cJ}\rightarrow
\pi^+\pi^-)=\sum_{i,j}^{\pi,p,\sigma}\int_0^1 dx \int_0^1 dy
\phi_i(x,\mu_F)T_{HJ}^{ij}(x,y,m_c^2)\phi_j(y,\mu_F),
\end{eqnarray}
where the renormalization scale $t_j$ is taken as the charm quark
mass and customarily identified with the factorization scale. The
hard-scattering amplitudes $T_{HJ}^{ij}(x,y,m_c^2)$ are expressed as
follows
\begin{eqnarray}
T_{H0}^{\pi\pi}(x,y,m_c^2)={\mathcal {N}_0 \alpha_s^2(m_c^2)\over 36
\sqrt{6}m_c^2} {x^2+(2-6 y) x+y (y+2)\over x(1-x)y(1-y)(x+y-2xy)^2},
\nonumber
\end{eqnarray}
\begin{eqnarray}
T_{H0}^{pp}(x,y,m_c^2)={\mathcal {N}_0 \alpha_s^2(m_c^2)\over 72
\sqrt{6}m_c^2}{\mu_\pi^2 \over m_c^2} {x^2 + (3 - 8 y) x + y (y +
3)\over x(1-x)y(1-y)(x+y-2xy)^2}, \nonumber
\end{eqnarray}
\begin{eqnarray}
T_{H0}^{p\sigma}(x,y,m_c^2)=-{\mathcal {N}_0 \alpha_s^2(m_c^2)\over
432 \sqrt{6}m_c^2}{\mu_\pi^2 \over m_c^2} {y^3+x (-2 y^2-5 y+5)
y+x^2 (8 y^2-8 y+1)\over x(1-x)y^2(1-y)^2(x+y-2xy)^3}, \nonumber
\end{eqnarray}
\begin{eqnarray}
T_{H0}^{\sigma p}(x,y,m_c^2)=-{\mathcal {N}_0 \alpha_s^2(m_c^2)\over
432 \sqrt{6}m_c^2}{\mu_\pi^2 \over m_c^2} {(1-2 y) x^3+y (8 y-5)
x^2+(5-8 y) y x+y^2\over x^2(1-x)^2y(1-y)(x+y-2xy)^3}, \nonumber
\end{eqnarray}
\begin{eqnarray}
T_{H0}^{\sigma\sigma}(x,y,m_c^2)&=&-{\mathcal {N}_0
\alpha_s^2(m_c^2)\over 2592 \sqrt{6}m_c^2}{\mu_\pi^2 \over m_c^2}
{1\over x^2(1-x)^2y^2(1-y)^2(x+y-2xy)^4}((24 y^3-36 y^2\nonumber \\
&&+14 y-1) x^5+(48 y^4-156 y^3+145 y^2-42 y+3) x^4+y (8 \ y^4-116
y^3\nonumber \\
&&+218 y^2-129 y+17) x^3+y (-12 y^4+89 y^3-97 y^2+24 y+4) \ x^2\nonumber \\
&&+y^2 (6 y^3-26 y^2+9 y+4) x-(y-3) y^4)
\end{eqnarray}
for $J=0$ and
\begin{eqnarray}
T_{H2}^{\pi\pi}(x,y,m_c^2)={\mathcal {N}_0 \alpha_s^2(m_c^2)\over 36
\sqrt{3}m_c^2} {x+y-x^2-y^2 \over x(1-x)y(1-y)(x+y-2xy)^2},
\nonumber
\end{eqnarray}
\begin{eqnarray}
T_{H2}^{pp}(x,y,m_c^2)=-{\mathcal {N}_0 \alpha_s^2(m_c^2)\over 72
\sqrt{3}m_c^2}{\mu_\pi^2 \over m_c^2} {(x-y)^2\over
x(1-x)y(1-y)(x+y-2xy)^2}, \nonumber
\end{eqnarray}
\begin{eqnarray}
T_{H2}^{p\sigma}(x,y,m_c^2)={\mathcal {N}_0 \alpha_s^2(m_c^2)\over
432 \sqrt{3}m_c^2}{\mu_\pi^2 \over m_c^2} {(8 y^2-8 y+1) x^2+2 y (-4
y^2+2 y+1) x+y^2 (4 y-3)\over x(1-x)y^2(1-y)^2(x+y-2xy)^3},
\nonumber
\end{eqnarray}
\begin{eqnarray}
T_{H2}^{\sigma p}(x,y,m_c^2)={\mathcal {N}_0 \alpha_s^2(m_c^2)\over
432 \sqrt{3}m_c^2}{\mu_\pi^2 \over m_c^2} {(4 - 8 y) x^3 + (8 y^2 +
4 y - 3) x^2 + 2 (1 - 4 y) y x + y^2)\over
x^2(1-x)^2y(1-y)(x+y-2xy)^3}, \nonumber
\end{eqnarray}
\begin{eqnarray}
T_{H2}^{\sigma\sigma}(x,y,m_c^2)&=&{\mathcal {N}_0
\alpha_s^2(m_c^2)\over 2592 \sqrt{3}m_c^2}{\mu_\pi^2 \over m_c^2}
{1\over x^2(1-x)^2y^2(1-y)^2(x+y-2xy)^4}((24 y^3-36 y^2\nonumber \\
&&+14 y-1) x^5+(-144 y^4+228 y^3-119 y^2+30 y-3) x^4+(8 \ y^5+268
y^4\nonumber \\
&&-424 y^3+210 y^2-46 y+3) x^3+y (-12 y^4-175 y^3+242 y^2-84 \ y+10)
x^2\nonumber \\
&&+2 y^2 (3 y^3+23 y^2-27 y+5) x-y^3 (y^2+3 y-3))
\end{eqnarray}
for $J=2$.
\section{the wave functions of light mesons}

In the above calculation, the twist-2,3 wave functions and
distribution amplitudes of pion and kaon are the main
non-perturbative input parameters for mHSA and sHSA, respectively.
In this section, we will discuss them in detail. According to BHL
prescription\cite{BHL}, one can connect the equal-time wave function
in the rest frame and the light-cone wave function by equating the
off-shell propagator in the two frames. The wave function for
quark-antiquark systems at the infinite momentum frame can be got
from the harmonic oscillator model at the rest frame
\begin{eqnarray}
\Psi(x,\textbf{k}_\perp)\propto exp \left [-{1\over 8 \beta^2}\left
({\textbf{k}_\perp^2+m_1^2\over x}+{\textbf{k}_\perp^2+m_2^2\over
1-x}\right )\right ],
\end{eqnarray}
where $m_i$ and $\beta$ are the constitute quark mass and the
harmonic parameter, respectively. The distribution amplitude can be
obtained from the integration of wave function over the transverse
momentum
\begin{eqnarray}
\phi(x,\mu_F)=\int_{|\textbf{k}_\perp|<\mu_F}{d^2\textbf{k}_\perp\over
16 \pi^3}\Psi(x,\textbf{k}_\perp),
\end{eqnarray}
where $\mu_F$ is the ultraviolet cutoff.

Upon expansion over Gegenbauer polynomials, twist-2 wave functions
of pion and kaon with the transverse momentum dependence can be
characterized as
\begin{eqnarray}
\Psi_\pi^\pi(x,\textbf{k}_\perp)=A_\pi^\pi\left[1+B_\pi^\pi
C_2^{3/2}(2x-1)+C_\pi^\pi C_4^{3/2}(2x-1)\right]exp \left
[-{\textbf{k}_\perp^2+m_q^2\over 8\beta_\pi^2 x(1-x )} \right ]
\end{eqnarray}
and
\begin{eqnarray}
\Psi_K^K(x,\textbf{k}_\perp)&=&A_K^K\left[1+B_K^K
C_1^{3/2}(2x-1)+C_K^K
C_2^{3/2}(2x-1)\right]\nonumber \\
&& \times exp \left [-{1\over
8\beta_K^2}\left({\textbf{k}_\perp^2+m_q^2\over
 x}+{\textbf{k}_\perp^2+m_s^2\over
 1-x}\right) \right ],
\end{eqnarray}
where $C_n^{3/2}$ are Gegenbauer polynomials and $q$ means light
quark $u$ or $d$. For the SU(2) isotopic symmetry, the odd expansion
terms do not appear in the pion wave functions. On the contrary, the
odd expansion terms are not zero in the Kaon wave functions for
SU(3)-flavor symmetry breaking. Estimates of first two Gegenbauer
moments for twist-3 distribution amplitudes are more uncertain than
that of leading twist distribution amplitudes. To simplify the
following numerical analysis, we take twist-3 wave functions as
\begin{eqnarray}
\Psi_p^\pi(x,\textbf{k}_\perp)={A_p^\pi\over x(1-x)}exp \left
[-{\textbf{k}_\perp^2+m_q^2\over 8\beta_\pi^2 x(1-x )} \right
],~\Psi_\sigma^\pi(x,\textbf{k}_\perp)=A_\sigma^\pi exp \left
[-{\textbf{k}_\perp^2+m_q^2\over 8\beta_\pi^2 x(1-x )} \right ]
\end{eqnarray}
and
\begin{eqnarray}
&&\Psi_p^K(x,\textbf{k}_\perp)={A_p^K\over x(1-x)} exp \left
[-{1\over 8\beta_K^2}\left({\textbf{k}_\perp^2+m_q^2\over
 x}+{\textbf{k}_\perp^2+m_s^2\over
 1-x}\right) \right ],\nonumber \\
&&\Psi_\sigma^K(x,\textbf{k}_\perp)=A_\sigma^K exp \left [-{1\over
8\beta_K^2}\left({\textbf{k}_\perp^2+m_q^2\over
 x}+{\textbf{k}_\perp^2+m_s^2\over
 1-x}\right) \right ]
\end{eqnarray}
for pion and kaon, respectively.

Substituting Eq.(24)-Eq.(27) into Eq.(23), the distribution
amplitudes of pion and kaon are written as
\begin{eqnarray}
\phi_\pi^\pi(x)={A_\pi^\pi \beta_\pi^2\over 2 \pi^2}
x(1-x)\left[1+B_\pi^\pi C_2^{3/2}(2x-1)+C_\pi^\pi
C_4^{3/2}(2x-1)\right]exp \left [-{m_q^2\over 8\beta_\pi^2 x(1-x )}
\right],
\end{eqnarray}
\begin{eqnarray}
\phi_K^K(x)&=&{A_K^K \beta_K^2\over 2 \pi^2}x(1-x)\left[1+B_i^K
C_1^{3/2}(2x-1)+C_i^K C_2^{3/2}(2x-1)\right]\nonumber \\
&&\times exp \left [-{(1-x)m_q^2+x m_s^2\over 8\beta_K^2 x(1-x)}
\right ],
\end{eqnarray}
for twist-2 distribution amplitudes and
\begin{eqnarray}
\phi_p^\pi(x)={A_p^\pi \beta_\pi^2 \over 2 \pi^2}exp \left
[-{m_q^2\over 8\beta_\pi^2 x(1-x )} \right
],~\phi_\sigma^\pi(x)={A_\sigma^\pi \beta_\pi^2 \over 2 \pi^2}exp
\left [-{m_q^2\over 8\beta_\pi^2 x(1-x )} \right ],
\end{eqnarray}
\begin{eqnarray}
\phi_p^K(x)={A_p^K \beta_K^2\over 2\pi^2)}exp \left [-{(1-x)m_q^2+x
m_s^2\over 8\beta_K^2 x(1-x)} \right ],~\phi_\sigma^K(x)={A_\sigma^K
\beta_K^2\over 2\pi^2)}exp \left [-{(1-x)m_q^2+x m_s^2\over
8\beta_K^2 x(1-x)} \right ]
\end{eqnarray}
for twist-3 distribution amplitudes. With the help of the above
distribution amplitudes from BHL prescription, the endpoint problem
can be cured in the standard HSA since the exponential suppression
appears in $x=0$ and $x=1$ point.

For definiteness, we take the conventional values for the constitute
quark masses: $m_q = 0.30 GeV$ and $m_s = 0.45 GeV$. The parameters,
$A_i^j$, $B_j^j$, $C_j^j$ and $\beta_j$
$(i=\pi,K,p,\sigma;~j=\pi,K)$ can be determined by some constraints
on the general properties of the light mesons wave functions. In the
pion and kaon case, the harmonic parameters $\beta_\pi$ and
$\beta_K$ are obtained by the constraints $<\textbf{k}_\perp^2>_K
\approx <\textbf{k}_\perp^2>_\pi\approx (0.356 GeV)^2$, which are
the average values of the transverse momentum square defined as
\begin{eqnarray}
<\textbf{k}_\perp^2>_i={f_i^2\over 24}\int dx {d^2
\textbf{k}_\perp\over 16
\pi^3}|\textbf{k}_\perp^2||\Psi_i(x,\textbf{k}_\perp)|^2 /
P^i_{q\bar{q}}
\end{eqnarray}
with $i=\pi,K$ and $\Psi_i$ stand for twist-2 wave functions
$\Psi_\pi^\pi$ and $\Psi_K^K$. The decay constants are taken as
$f_\pi=0.132~GeV$ for the pion and $f_K=0.160~GeV$ for the kaon. The
probability of finding the $q\bar{q}$ leading-twist Fock state in a
pion or kaon should be not larger than unity,
\begin{eqnarray}
P^i_{q\bar{q}}={f_i^2\over 24}\int dx {d^2 \textbf{k}_\perp\over 16
\pi^3}|\Psi_i(x,\textbf{k}_\perp)|^2\leq 1.
\end{eqnarray}

The others coefficients are extracted by the normalization condition
\begin{eqnarray}
\int dx {d^2 \textbf{k}_\perp\over 16 \pi^3}
\Psi_i^j(x,\textbf{k}_\perp)=1
\end{eqnarray}
with $(i=\pi,K,p,\sigma;~j=\pi,K)$ and first two Gegenbauer moments
of twist-2 distribution amplitudes $a_i^j$ ($i=2,4$ for $j=\pi$;
$i=1,2$ for $j=K$.). The coefficients $a_i^j$ at some reference
scale $\mu_F$ are nonperturbative quantities and have to be
evaluated using a nonperturbative technique or must be extracted
from experiment. It turns out that the determination of a¼ 2 receive
large errors, whether by direct calculations using QCD sum
rules\cite{DAmoment1} or by analysis of experimental data on the
pion electromagnetic and transition form factors\cite{DAmoment2}.
Totally, the averages of the second moment are probably
\begin{eqnarray}
a_2^\pi(1~GeV)=0.25\pm0.15,~~a_2^K(1~GeV)=0.25\pm0.15,
\end{eqnarray}
in Ref.\cite{DA2}, including radiative corrections to the sum rules.

The numerical value of the first moment $a_1^K$ was the subject of
significant controversy until recently. The existing estimates are
all obtained using different versions of QCD sum rules
\cite{DAmoment4,DAmoment5,DAmoment6,DAmoment7} and yield an average
\cite{DA2}
\begin{eqnarray}
a_1^K(1~GeV)=0.06\pm0.03.
\end{eqnarray}
Estimates of yet higher-order Gegenbauer moments are rather
uncertain. The fourth Gegenbauer moment of the pion twist-2
distribution amplitude \cite{piong1} was constrained
\begin{eqnarray}
a_4^\pi(1~GeV)=0.04\pm0.11,
\end{eqnarray}
which is consistent with the results from the light-cone sum rule
calculations of the transition form factor $F_{\pi \gamma\gamma^*}$
in Refs. \cite{piong2,piong3,piong4}.

\begin{table}
\caption{The parameters  of twist-2,3 wave functions for the pion
and kaon mesons in Eq.(24)-Eq.(31). The dimensions of harmonic
parameters $\beta_{\pi,K}$ and normalization coefficients $A_i^j$
$(i=\pi,K,p,\sigma;j=\pi,K.)$ are $GeV$ and $(GeV)^{-2}$,
respectively. The others parameters is dimensionless.}
\begin{ruledtabular}
\begin{tabular}{cccccccc}
$\pi$ & $\beta_\pi$ & $P_{q\bar{q}^\pi}$ & $A_\pi^\pi$
 & $B_\pi^\pi$ & $C_\pi^\pi$&$A_p^\pi$&$A_\sigma^\pi$   \\
\hline upper limit &0.512&0.270&672.28&0.628&0.354&106.92&574.47\\
      central value &0.461&0.249&849.18&0.469&0.213&140.88&747.55\\
       lower limit &0.418&0.259&1034.30&0.259&0.023&184.29&965.85\\
\hline $K$ &$\beta_K$ &$P_{q\bar{q}}^K$ &$A_K^K$ & $B_K^K$ & $C_K^K$ &$A_p^K$&$A_\sigma^K$ \\
\hline upper limit &0.461&0.492&1108.83&0.253&0.618&170.59&883.49\\
       central value &0.442&0.422&1196.66&0.215&0.477&193.81&998.32\\
       lower limit &0.417&0.398&1353.56&0.175&0.323&232.44&1188.23\\
\end{tabular}
\end{ruledtabular}
\end{table}

According to QCD evolution of the wave function, the coefficients
$a_i^j$ at a factorization scale $\mu_F$ can be expressed as
$a_i^j(\mu_F^2)=a_i^j(\mu_0^2)(\alpha_s(\mu_F^2)/\alpha_s(\mu_0^2))^{\gamma_i}$.
$a_i^j(\mu_0^2)$ are a non-perturbative coefficients, $\mu_0$ is a
typical hadronic scale, $0.5\leq\mu_0\leq1$ GeV, and $\gamma_i$ are
the anomalous dimensions. In this work, a reasonable factorization
scale should be chosen as $\mu_F=m_c$, $1.35\leq m_c\leq 1.8$ GeV.
In Ref.\cite{DA2}, the values of the coefficients $a_i^j$ at
$\mu_F=1$ GeV and $\mu_F=2$ GeV are listed in Table.3. For example,
$a_1^K(1 GeV)=0.06\pm0.03$ and $a_1^K(2 GeV)=0.05\pm0.02$. By
analyzing these data, we find that it is feasible to choose
Gegenbauer moments $a_i^j$ at $\mu_F=1$ GeV in our calculation.

Taking account of the above Gegenbauer moments for the pion and kaon
twist-2 distribution amplitudes, we figure out harmonic parameters
$\beta_{\pi,K}$, probabilities of finding the $q\bar{q}$
leading-twist Fock state $P_{q\bar{q}}^{\pi,K}$ and Gegenbauer
coefficients $A_i^j$, $B_j^j$, $C_j^j$
$(i=\pi,K,p,\sigma;~j=\pi,K)$. Those values are list in Table.II.
According to uncertainties of twist-2 Gegenbauer moments, the
parameters of the pion and kaon are given in three parts: upper
limit, central value and lower limit. Since
$P_{q\bar{q}}^{\pi}\leq0.270$ and $P_{q\bar{q}}^{K}\leq0.492$ are
much smaller than unity, higher twist and higher Fock states are
important components of the pion and kaon.

In the mHSA, the convolutions of wave functions and hard- scattering
amplitudes are presented in transverse configuration b-space. We
need to define wave function in b-space by Fourier transformation
\begin{eqnarray}
\Psi^i(x,\textbf{b})=\int {d^2 \textbf{k}_\perp \over (2\pi)^2}
\Psi^i(x,\textbf{k}_\perp) e^{-i \textbf{k}_\perp\cdot
\textbf{b}}=4\pi \phi^i(x)exp\left [ -2 \beta_i^2 x(1-x)b^2 \right
],
\end{eqnarray}
where $\Psi^i$ and $\phi^i$ stand for wave function $\Psi_j^i$ and
distribution amplitude $\phi_j^i$ $(i=\pi,K,p,\sigma;~j=\pi,K)$,
respectively. One may observe that wave functions in the b-space are
also highly suppressed in the endpoint region. Such feature is
necessary to suppress the endpoint singularity coming from the
hard-scattering amplitudes and then to derive a more reasonable
results.

\section{numerical analysis}

In our calculations for the decay ratio of $\chi_{cJ}$ to light
pseudoscalar pairs, the partial of P-wave function at the origin
$R_p^{'}(0)$ is also an important nonperturbative input parameter.
It is shown that this parameter is a function of the charm-quark
mass $m_c$ both in the well-know quarkonium potential models
\cite{pmod1,pmod2} and in the global fit of charmonium parameters
\cite{fitcp}. To obtain its expression relative to the charm-quark
mass, which is consisted with our approach, the decay width of the
$\chi_{c0}$ annihilating into two photon need to be calculated by
the same approach. With the help of
Refs.\cite{fitcp,tpQCD1,tpQCD2,tpQCD3}, we obtain
\begin{eqnarray}
\Gamma[\chi_{c0}\rightarrow\gamma\gamma]=27 Q_c^4 \alpha_{em}^2
{|R_p^{'}(0)|^2\over m_c^4}\left [ 1+({\pi^2\over 3}-{28\over
9}){\alpha_s(m_c^2)\over \pi}\right ],
\end{eqnarray}
where the one-loop QCD radiative correction is included, $Q_c={2
\over 3}$ is the charge of the c-quark and $\alpha_{em}={1\over
137}$ is the electromagnetic coupling constant.

The running coupling constant $\alpha_s(Q^2)$ up to
next-to-leading-log is written as
\begin{eqnarray}
{\alpha_s(Q^2)\over \pi}={1\over \beta_1
\ln(Q^2/\Lambda_{QCD}^2)}-{\beta_2\over \beta_1^3} {\ln
\ln(Q^2/\Lambda_{QCD}^2)\over \ln^2(Q^2/\Lambda_{QCD}^2)}
\end{eqnarray}
with $\beta_1=(33-2 n_f)/12$ and $\beta_2=(153-19 n_f)/24$. Here we
take quark-flavor number $n_f=4$ and the QCD scale
$\Lambda_{QCD}=0.25~GeV$.

The decay width of the $\chi_{c0}$ annihilating into two photon can
be obtained from Refs.\cite{Cleo08,pdg2008}. Using the above data
and formulas, the relation between $R_p^{'}(0)$ and $m_c$ is shown
in Fig.2 with uncertainty of the $\chi_{c0}$ two-photon decay width.
The solid curve comes from taking central value
$\Gamma[\chi_{c0}\rightarrow \gamma\gamma]=2.37$ keV. The dot-dashed
and dashed curves are given by taking upper limit
$\Gamma[\chi_{c0}\rightarrow \gamma\gamma]=2.71$ keV and lower limit
$\Gamma[\chi_{c0}\rightarrow \gamma\gamma]=2.03$ keV, respectively.
We find that uncertainty with different $\chi_{c0}$ two-photon decay
width is less than $10\%$. So we will take the result of the central
value in the following. The region of the charm-quark mass is
$m_c=1.35-1.8$ GeV from Ref.\cite{octet}. Comparing our $R_p^{'}(0)$
with values of Ref.\cite{octet}, there are some differences that our
value is less than one of Ref.\cite{octet} as $m_c=1.35$ GeV and
vice versa as $m_c=1.8$ GeV.
\begin{figure}[ht]
\includegraphics[scale=0.7]{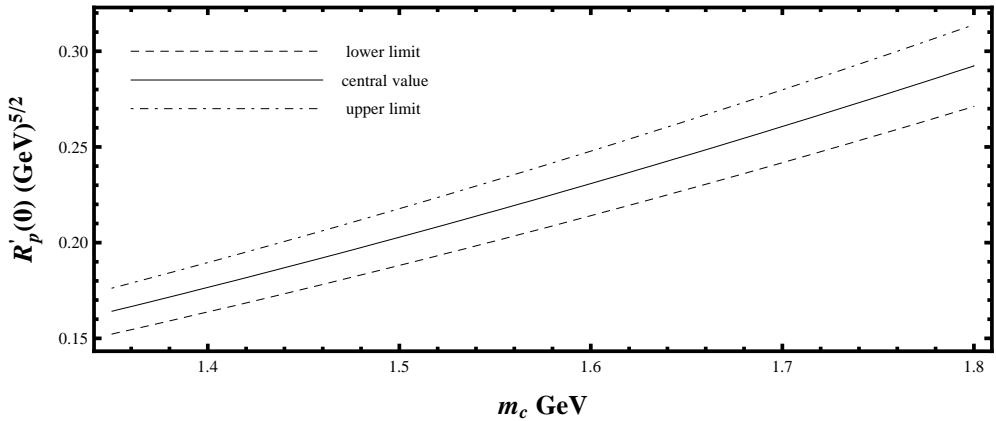}
\caption{Dependence of the partial of P-wave function at the origin
$R_p^{'}(0)$ for P-wave charmonium on the c-quark mass $m_c$ with
uncertainty of the $\chi_{c0}$ two-photon decay width.  }
\end{figure}

On the other hand, the chiral enhancing scales $\mu_\pi$ and
$\mu_K$, which are scales characterized by the chiral perturbation
theory, are important parameters which can affect contributions from
twist-3 parts sensitively. However, they are difficult to give
precise numbers as long as the current quark masses are not more
accurately known. To obtain reasonable numerical analysis with
acceptable estimates for the chiral enhancing scales, we take
$\mu_\pi(1~GeV)=1.5$ GeV and $\mu_K(1~GeV)=1.7$ GeV, which are
consistent with the results from pQCD application \cite{Ap1,Ap2} and
chiral perturbation theory \cite{Chtp}.

In the above discussion, nonperturbative input parameters appearing
in our calculation are obtained by a model wave function or
distribution amplitude, a fit experimental data and a reasonable
evaluation. Next, we will do numerical analysis for the charmonium
$\chi_{c0,2}$ decay rates into two pions or two kaons with the
charge case in the sHSA and in the mHSA, respectively.

\begin{figure}[ht]
\begin{center}
\setlength{\unitlength}{1cm}
\begin{minipage}[t]{7.5cm}
\begin{picture}(7.0,5.5)
\includegraphics*[scale=0.65,angle=0.]{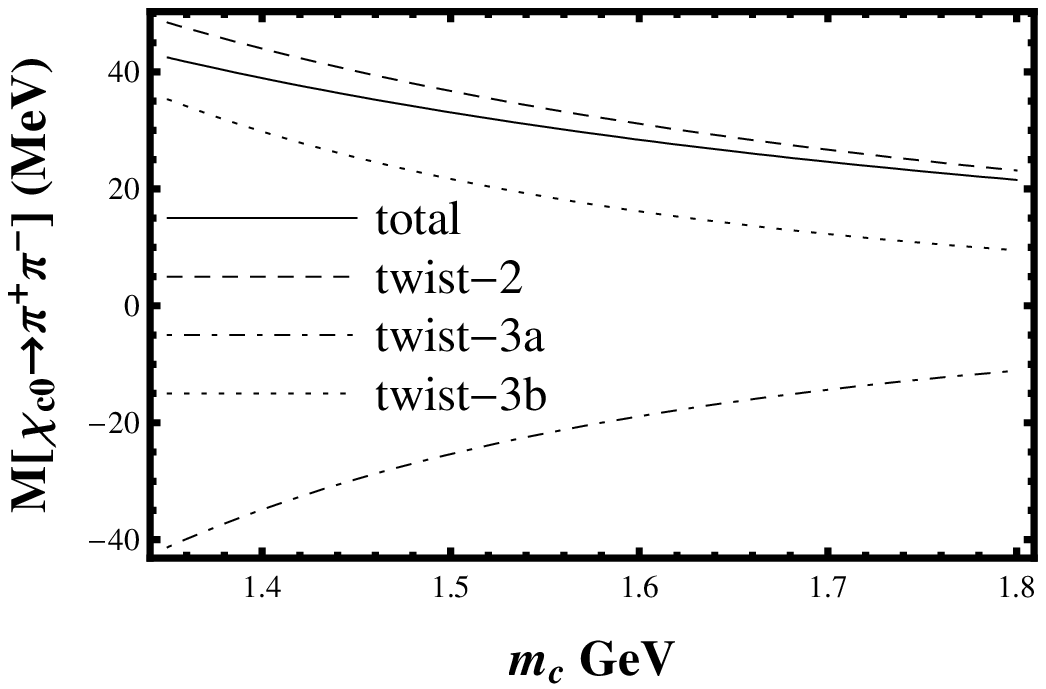}
\end{picture}\par
\end{minipage}
\hfill 
\begin{minipage}[t]{7.5cm}
\begin{picture}(7.0,5.5)
\includegraphics*[scale=0.65,angle=0.]{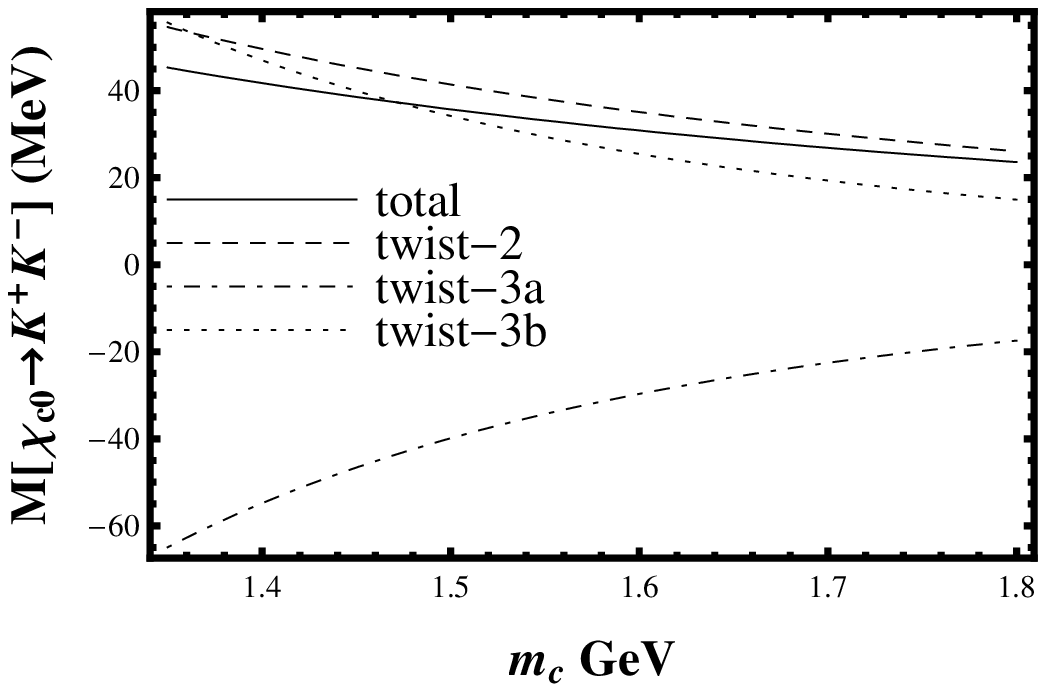}
\end{picture}\par
\end{minipage}
\vfill
\begin{minipage}[t]{7.5cm}
\begin{picture}(7.0,5.5)
\includegraphics*[scale=0.65,angle=0.]{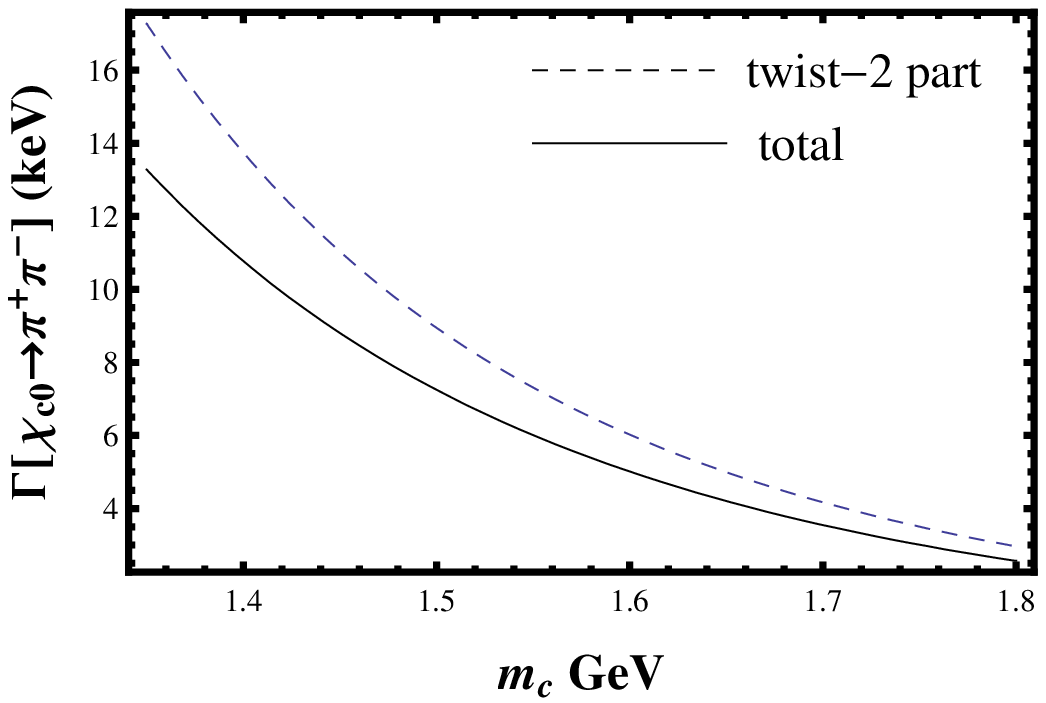}
\end{picture}\par
\end{minipage}
\hfill
\begin{minipage}[t]{7.5cm}
\begin{picture}(7.0,5.5)
\includegraphics*[scale=0.65,angle=0.]{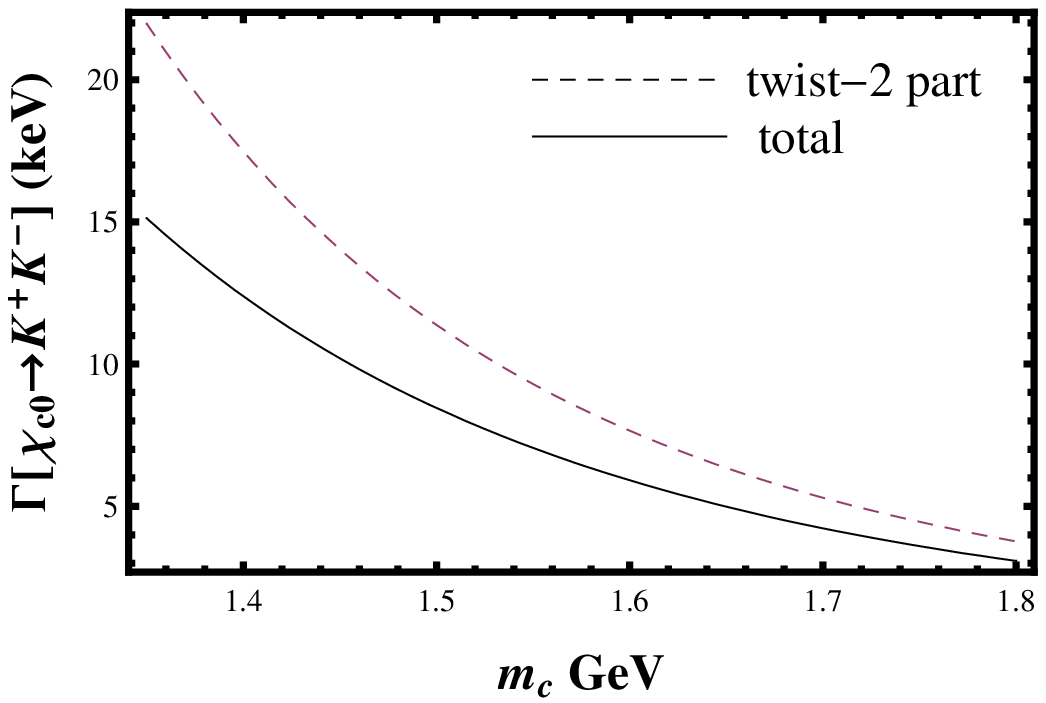}
\end{picture}\par
\end{minipage}
\vfill
\begin{minipage}[t]{7.5cm}
\begin{picture}(7.0,5.5)
\includegraphics*[scale=0.65,angle=0.]{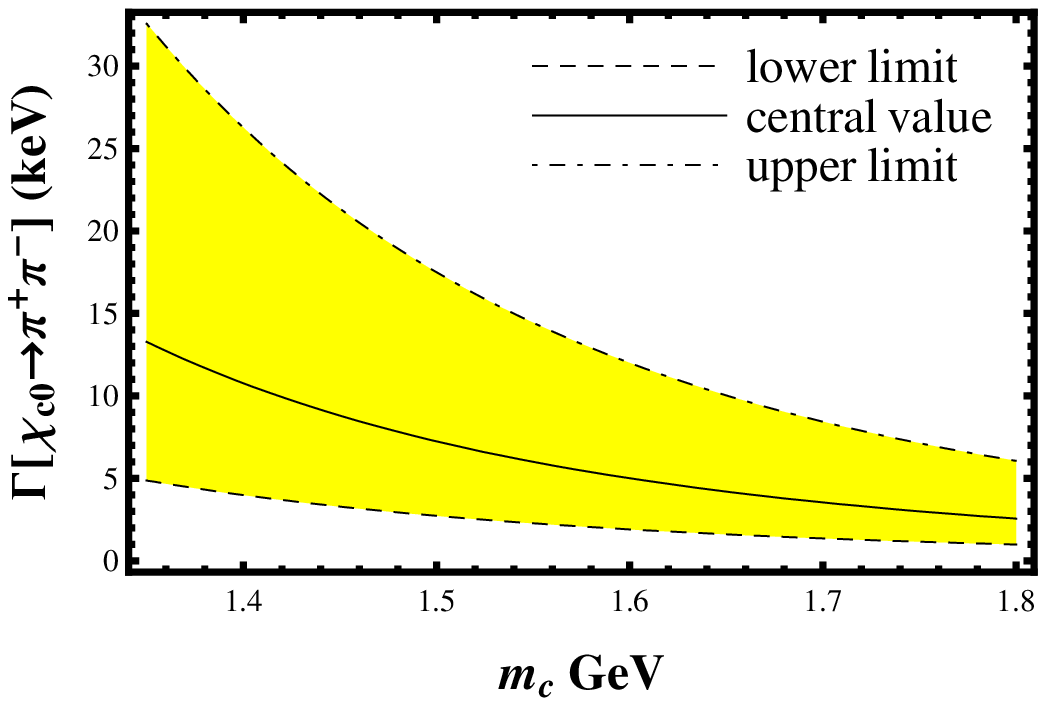}
\end{picture}\par
\end{minipage}
\hfill
\begin{minipage}[t]{7.5cm}
\begin{picture}(7.0,5.5)
\includegraphics*[scale=0.65,angle=0.]{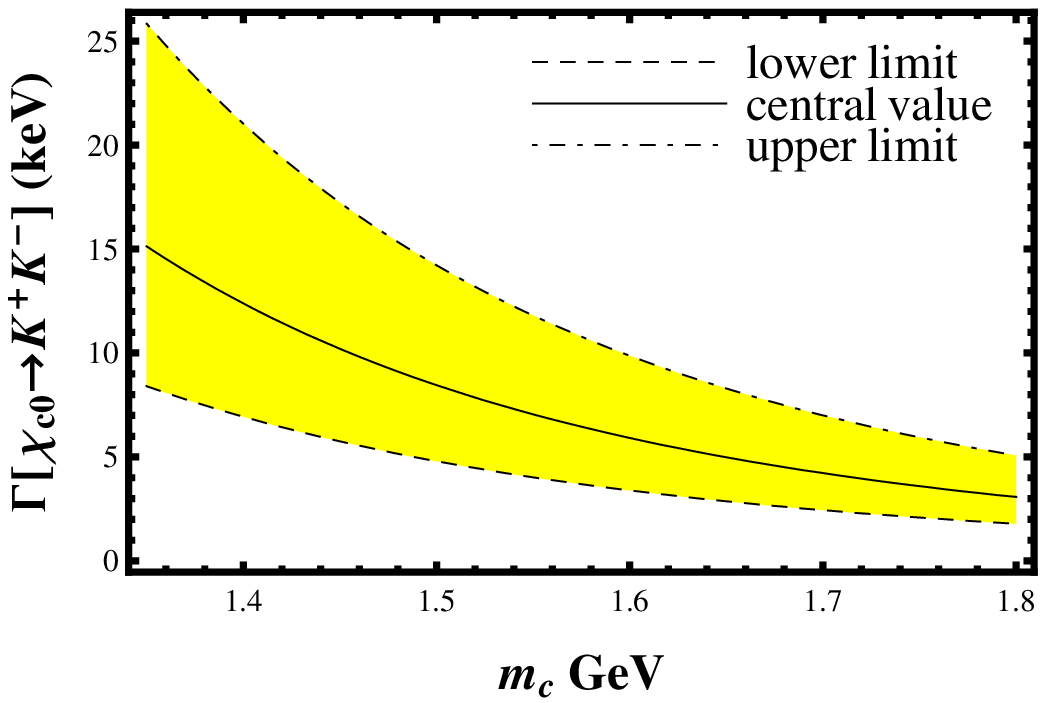}
\end{picture}\par
\end{minipage}
\caption{Dependence of the prediction for the $\chi_{c0}\rightarrow
\pi^+\pi^-$ and $K^+K^-$ decay widths on the c-quark mass $m_c$ with
contribution from twist-3 distribution amplitudes of the pion and
kaon meson in sHSA, respectively.}
\end{center}
\end{figure}
\begin{figure}[ht]
\begin{center}
\setlength{\unitlength}{1cm}
\begin{minipage}[t]{7.5cm}
\begin{picture}(7.0,5.5)
\includegraphics*[scale=0.65,angle=0.]{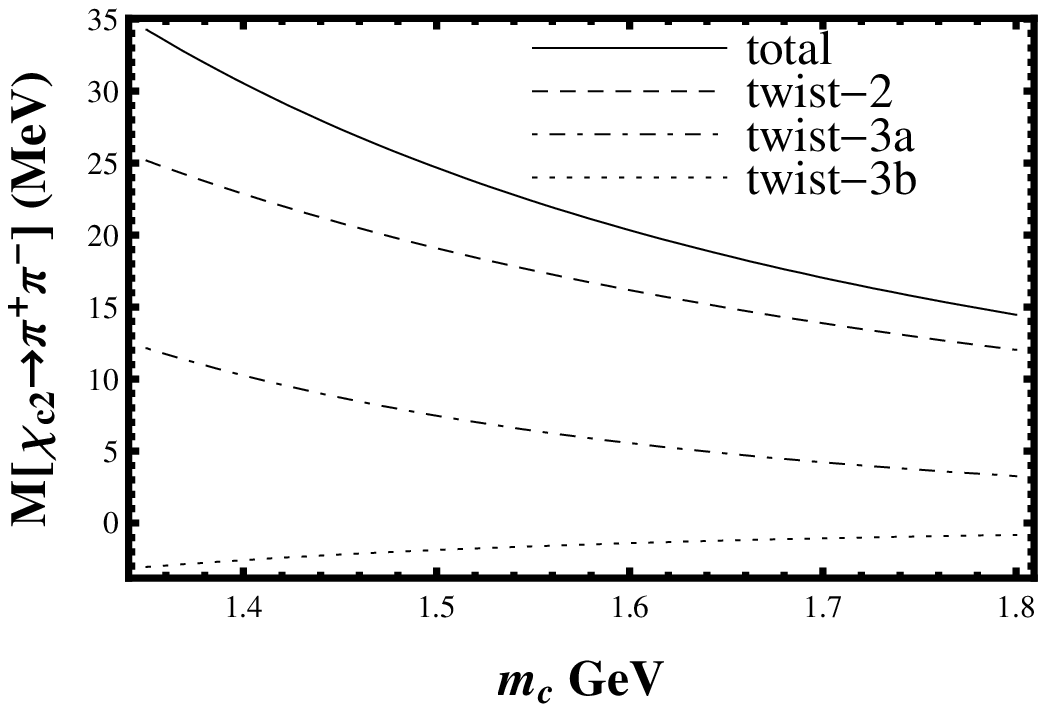}
\end{picture}\par
\end{minipage}
\hfill 
\begin{minipage}[t]{7.5cm}
\begin{picture}(7.0,5.5)
\includegraphics*[scale=0.65,angle=0.]{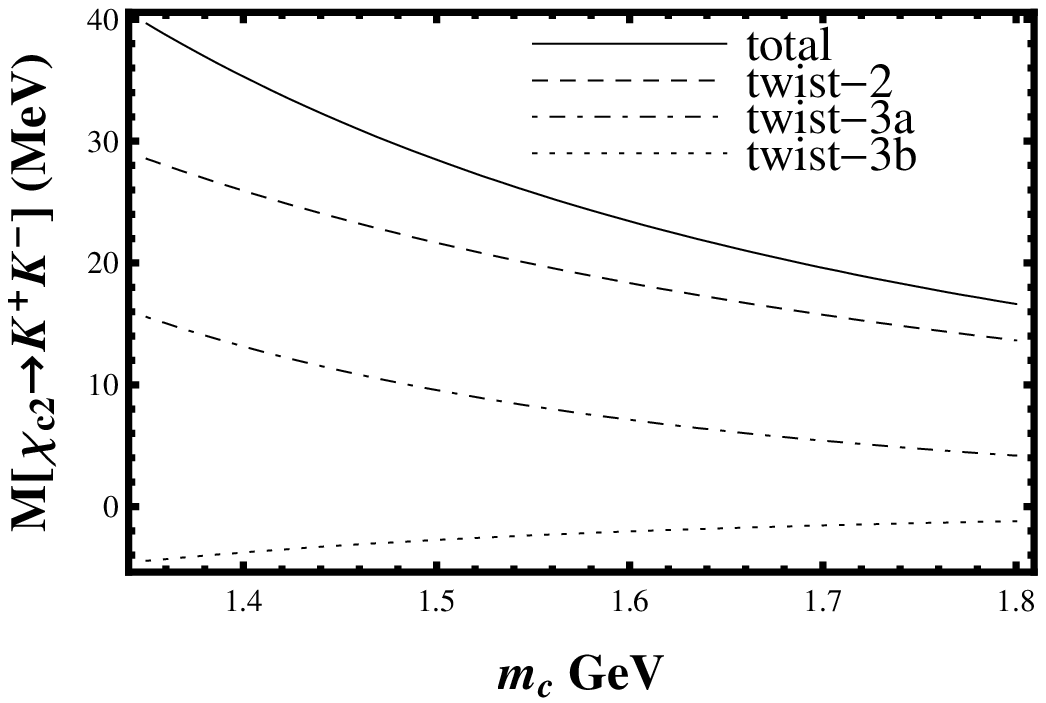}
\end{picture}\par
\end{minipage}
\vfill
\begin{minipage}[t]{7.5cm}
\begin{picture}(7.0,5.5)
\includegraphics*[scale=0.65,angle=0.]{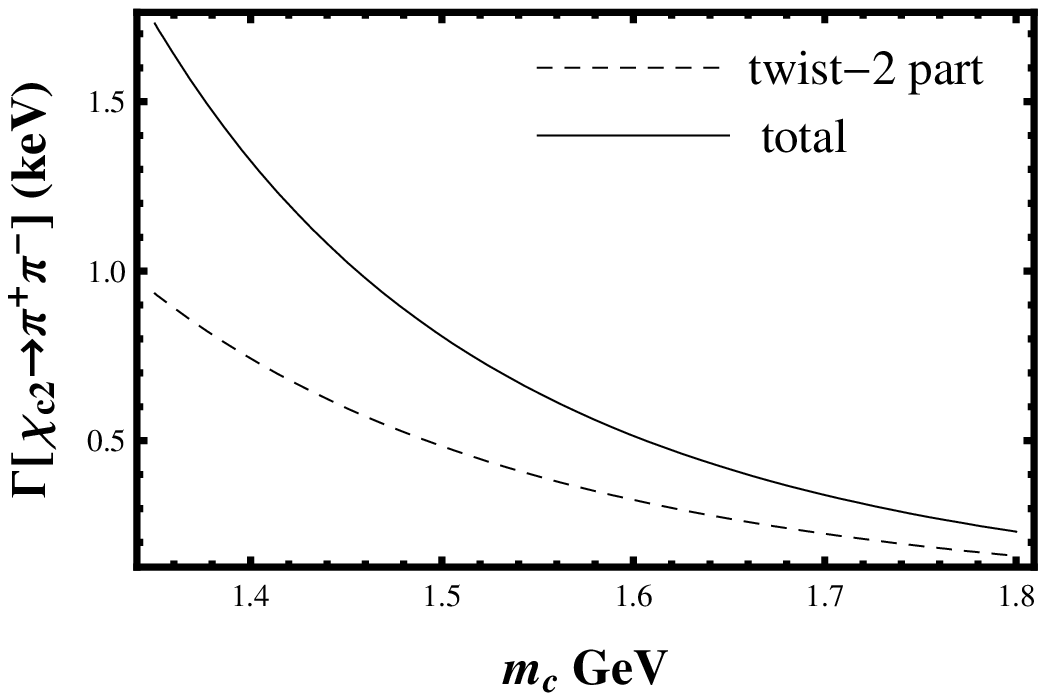}
\end{picture}\par
\end{minipage}
\hfill
\begin{minipage}[t]{7.5cm}
\begin{picture}(7.0,5.5)
\includegraphics*[scale=0.65,angle=0.]{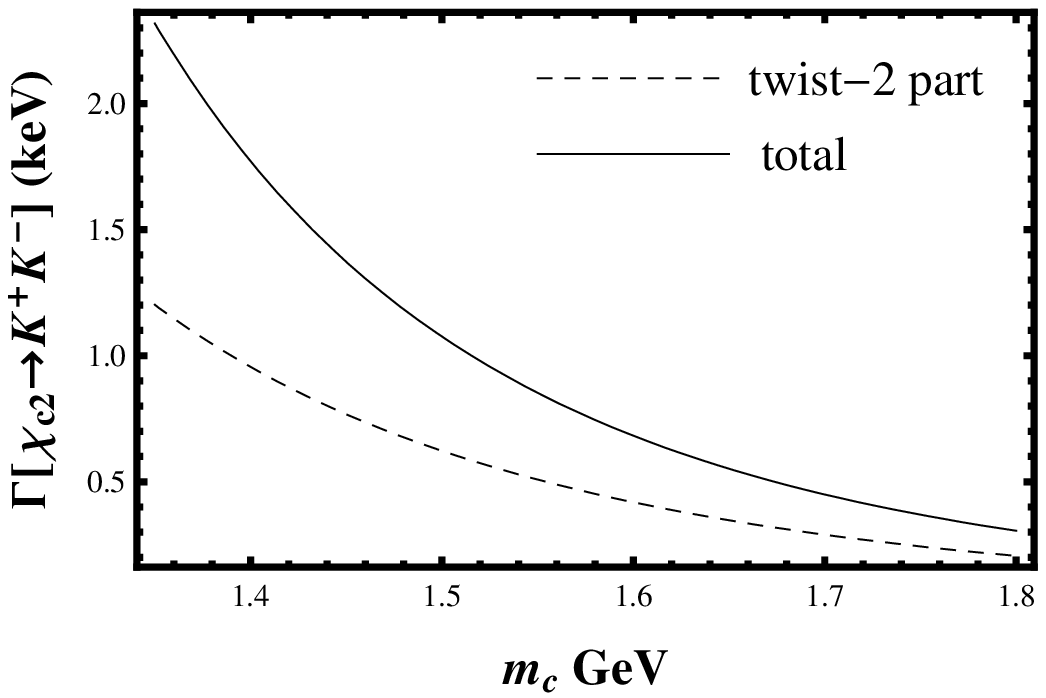}
\end{picture}\par
\end{minipage}
\vfill
\begin{minipage}[t]{7.5cm}
\begin{picture}(7.0,5.5)
\includegraphics*[scale=0.65,angle=0.]{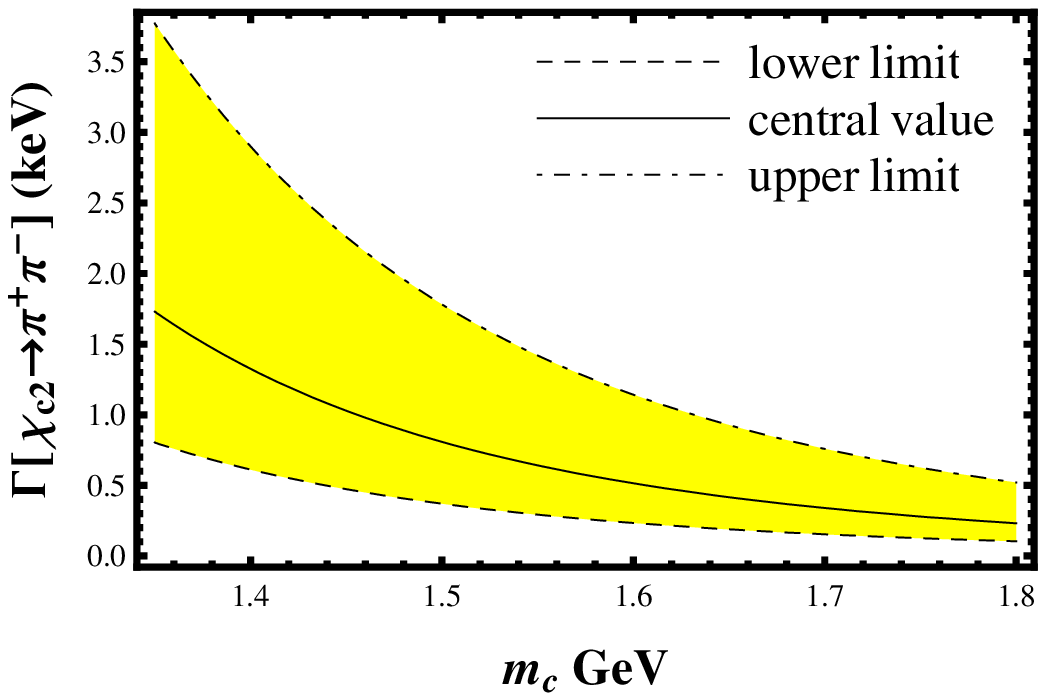}
\end{picture}\par
\end{minipage}
\hfill
\begin{minipage}[t]{7.5cm}
\begin{picture}(7.0,5.5)
\includegraphics*[scale=0.65,angle=0.]{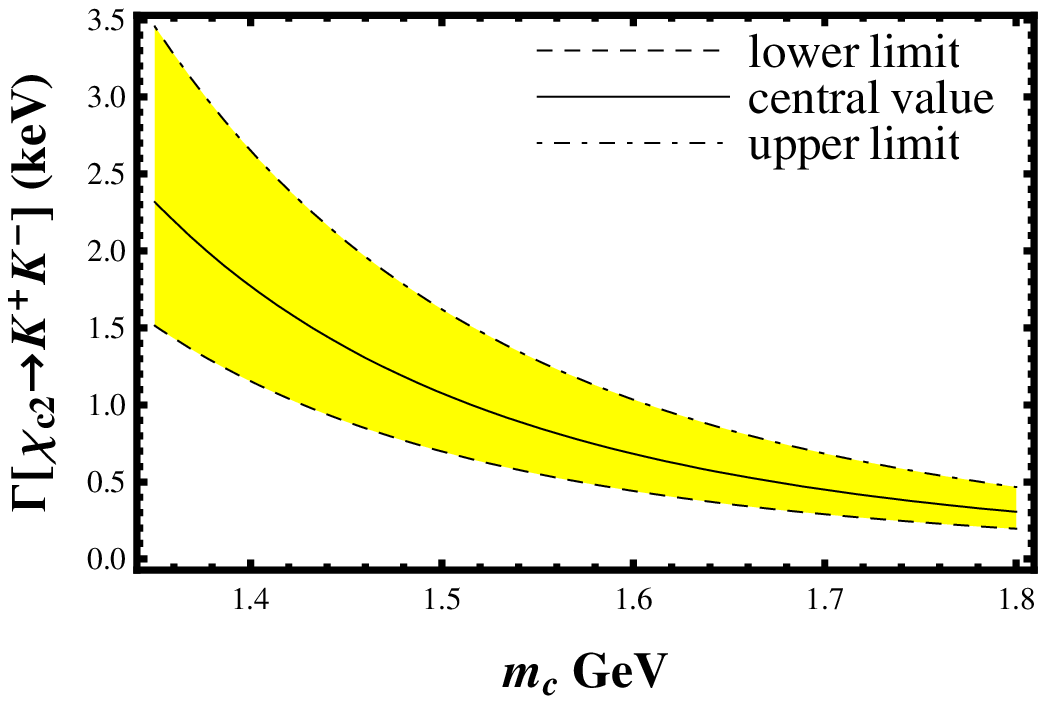}
\end{picture}\par
\end{minipage}
\caption{Dependence of the prediction for the $\chi_{c2}\rightarrow
\pi^+\pi^-$ and $K^+K^-$ decay widths on the c-quark mass $m_c$ with
contribution from twist-3 distribution amplitudes of the pion and
kaon meson in sHSA, respectively.}
\end{center}
\end{figure}
The decay widths of $\chi_{cJ}\rightarrow\pi^+\pi^-$ and $K^+K^-$
$(J=0,2)$ by the sHSA are shown in Fig.3 and Fig.4 with the c-quark
mass $m_c$ as a variable parameter. We take the central-value of
Table.II as the input parameters for the top two and middle two
figures. The top two figures are the decay amplitudes for two-pion
and two-kaon decays, where the dashed curve is the contribution from
twist-2 part, the dotted and dot-dashed are contributions from
twist-3 parts named a and b and the solid curve is the sum. The
twist-3a is the positive and the twist3b is the negative which are
corresponding to the positive and negative terms in Eq.(20) and
Eq.(21).

The large enhancement of the total amplitudes in the $\chi_2$
channel indicates that the twist-3 distribution amplitudes play
important pole. The results for the decay width are shown in the
middle two figures of Fig.4 where the solid curve is the total decay
width and the dashed curve is the for twist-2 part. In the $\chi_0$
channel, the corrections are not so large since the two
contributions from twist-3a and b parts have opposite sign and have
large cancelation. The results with the uncertainty of twist-2
distribution amplitude are shown in the bottom two figures, where
the solid curve is for the central value, the dashed and dot-dashed
curve are for lower limit and upper limit, respectively. We can see
the uncertain is very large which shows the sensitivity on the
distribution amplitude.

Comparing the results with experimental data
\cite{BES98,Belle05,pdg2008}, which are list in Table.III, we see
that the decay widths of $\chi_0$ channel are smaller than
experimental data for all variable $m_c$ and the decay widths of the
$\chi_2$ channel are in agreement with the experimental data in the
region $m_c<1.5$ GeV.
\begin{table}
\caption{The decay widths for $\chi_{cJ}\rightarrow \pi^+\pi^-$ and
$K^+K^-$ $(J=0,2)$ from experimental data. The BES results are
evaluated with the BES result for the total width. In the other
cases the PDG average values for the total widths are used.}
\begin{ruledtabular}
\begin{tabular}{cccc}
&PDG\cite{pdg2008}&BES\cite{BES98}&Belle\cite{Belle05} \\
\hline $\Gamma[\chi_{c0}\rightarrow \pi^+\pi^-]$[kev]&$50\pm8$&$67\pm36$&$60\pm21$\\
       $\Gamma[\chi_{c0}\rightarrow K^+K^-]$[kev]&$60\pm10$&$81\pm45$&$57\pm19$\\
       $\Gamma[\chi_{c2}\rightarrow \pi^+\pi^-]$[kev]&$2.8\pm0.5$&$3.0\pm1.0$&$3.4\pm1.3$\\
       $\Gamma[\chi_{c2}\rightarrow K^+K^-]$[kev]&$1.5\pm0.4$&$1.6\pm0.7$&$2\pm0.9$\\
\end{tabular}
\end{ruledtabular}
\end{table}

In Fig.5, we show the curves of the widths of the $\chi_{c0,2}$ into
two pions or two kaons on the c-quark mass mc by the mHSA method.
The solid curve is the decay width where only the twist-2
contribution is considered and the parameters are taken as the
central values. The shadow is the total decay width where the
uncertainty of twist-2 Gegenbauer moments are considered. Here we
see that the shade region is very narrow which means the
un-sensitivity on the twist-2 distribution amplitude of the light
mesons. The decay widthes including twist-3 corrections are improved
remarkably. Detailedly, the predictions for the decay widths of
$\chi_2\rightarrow \pi^+\pi^-$ and $\chi_0\rightarrow K^+K^-$ are
comparable with experimental data in the region $m_c \in(1.4,1.6)$
GeV and $m_c \in(1.35,1.6)$ GeV, respectively. This is very
different with the sHSA method which suggests the necessary of the
mHSA method.

\begin{figure}[hb]
\begin{center}
\setlength{\unitlength}{1cm}
\begin{minipage}[t]{7.5cm}
\begin{picture}(7.0,5.5)
\includegraphics*[scale=0.65,angle=0.]{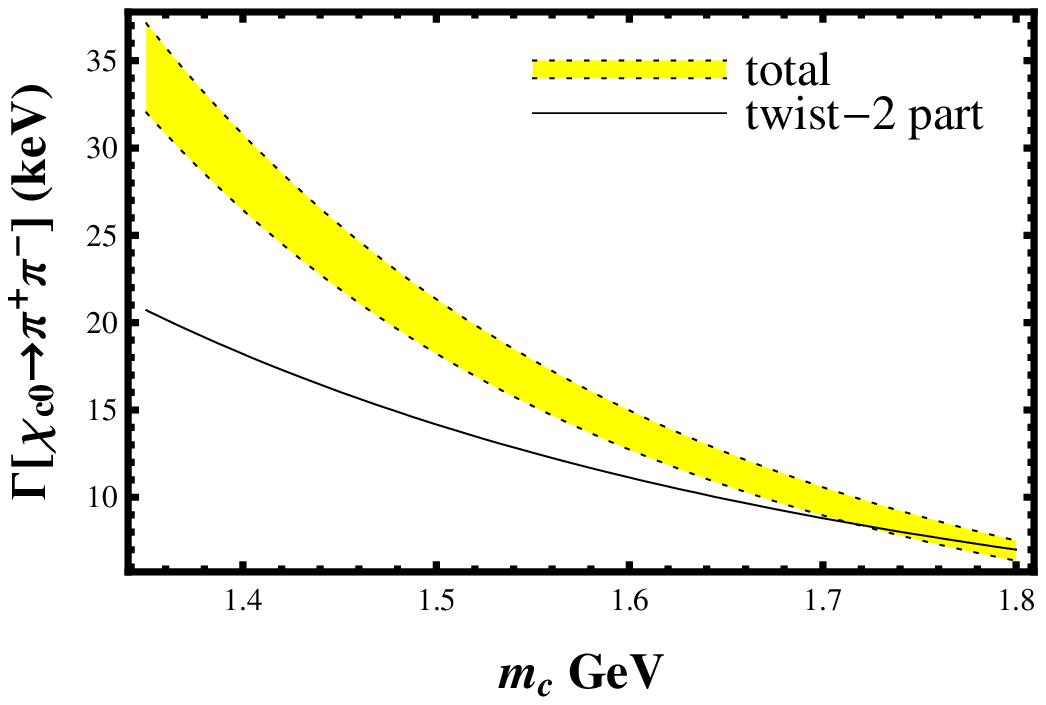}
\end{picture}\par
\end{minipage}
\hfill 
\begin{minipage}[t]{7.5cm}
\begin{picture}(7.0,5.5)
\includegraphics*[scale=0.65,angle=0.]{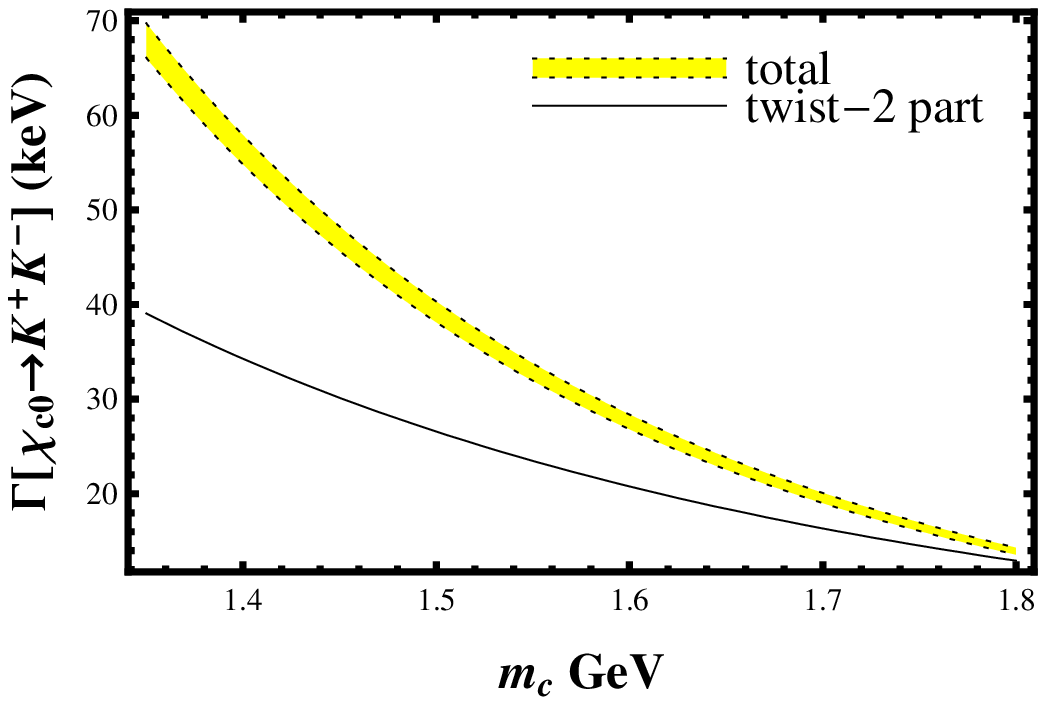}
\end{picture}\par
\end{minipage}
\vfill
\begin{minipage}[t]{7.5cm}
\begin{picture}(7.0,5.5)
\includegraphics*[scale=0.65,angle=0.]{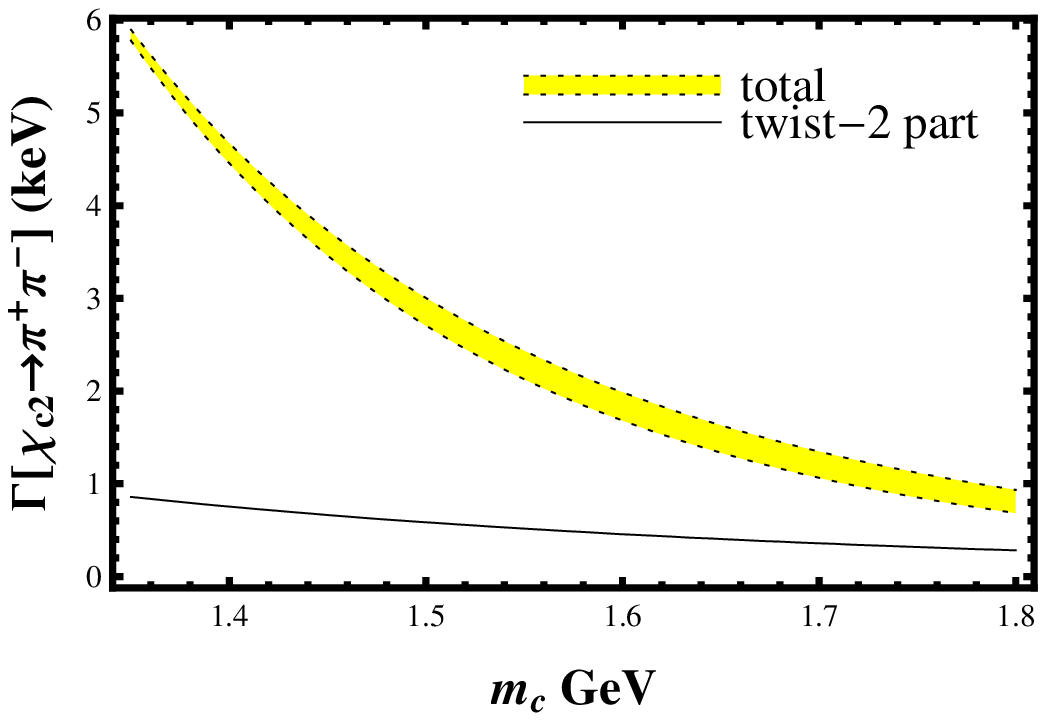}
\end{picture}\par
\end{minipage}
\hfill
\begin{minipage}[t]{7.5cm}
\begin{picture}(7.0,5.5)
\includegraphics*[scale=0.65,angle=0.]{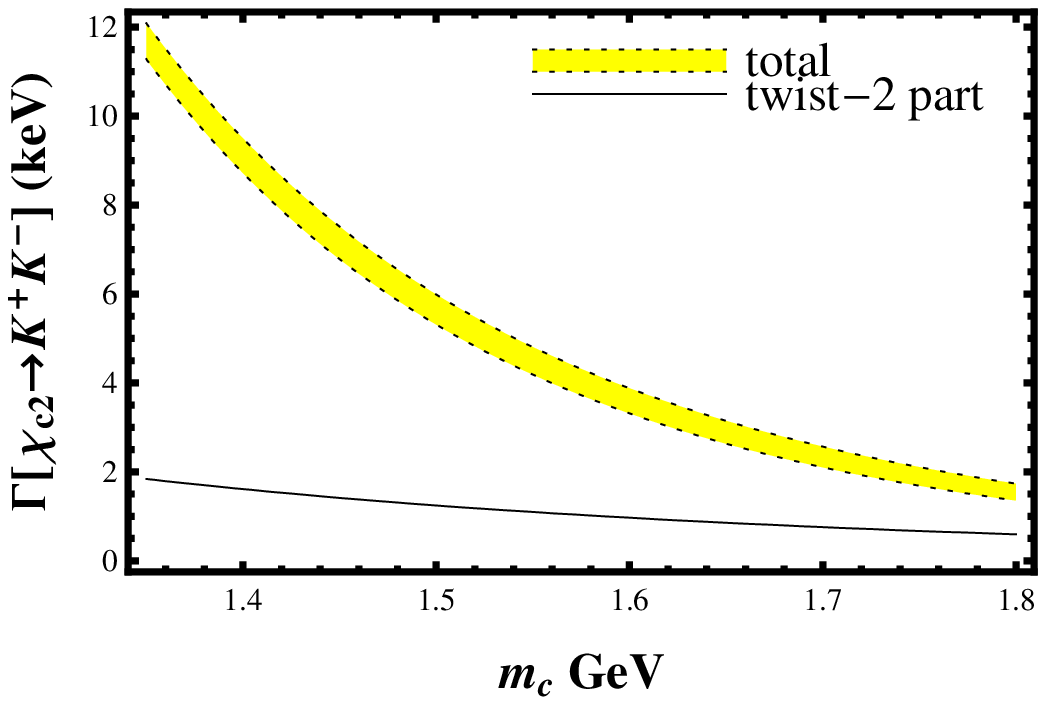}
\end{picture}\par
\end{minipage}
\caption{Dependence of the prediction for the $\chi_{cJ}\rightarrow
\pi^+\pi^-$ and $K^+K^-$ (J=0,2) decay widths on the c-quark mass
$m_c$ with contribution from twist-3 distribution amplitudes of the
pion and kaon meson in mHSA, respectively.}
\end{center}
\end{figure}

\section{conclusion}

In this paper, we presented a detailed analysis of $\chi_{c0,2}$
decays into two pions and two kaons including the twist-3
contribution within the framework of the sHSA and mHSA methods. In
the sHSA, the end-point problem is overcame by using BHL
prescription where a exponential suppression is introduced in the
expression of hadronic wave functions or distribution amplitudes.
The uncertainty of the results on the twist-2 Gegenbauer moments for
the pion and kaon is analyzed and is rather small in the mHSA
method. The results indicate the larger contributions from twist-3
distribution amplitude which have not been analyzed before. And both
the decay widthes of $\chi_{c0,2}$ to $\pi^+\pi^-$ and $K^+K^-$ are
found to be comparable with the experimental data in the region $m_c
\in(1.35,1.8)$ GeV when including twist-3 correction in the mHSA.

\appendix
\section{the coefficients of hard-scattering amplitudes}

In this Appendix we present the explicit expression of the
coefficients $C_i^I(J)(I=pp,p\sigma,\sigma p,\sigma\sigma)$
appearing in the Eq.(14) for hard-scattering amplitudes.
\begin{eqnarray}
&&C_1^{pp}(0)=5, ~~~C_0^{pp}(0)=2 m_c^2 (x^2+2 (y-1) x+(y-2) y), \\
&&C_1^{pp}(2)=2, ~~~C_0^{pp}(2)=-4 m_c^2 (x^2-4 y x+x+y^2+y)
\end{eqnarray}
for hard-scattering amplitude
$T_{HJ}^{pp}(x,y,\textbf{k}_1,\textbf{k}_2)$ with $J=0,2$.
\begin{eqnarray}
&&C_3^{p\sigma}(0)=5 m_c^2, ~~~C_2^{p\sigma}(0)=4 m_c^4 (4 x^2+(4 y-6) x-2 y-3),\nonumber\\
&&C_1^{p\sigma}(0)=-4 m_c^6 (11 x^2+2 (7 y-9) x+y (3 y-10)), \nonumber\\
&&C_0^{p\sigma}(0)=-16 m_c^8 (2 x^4+(6 y-7) x^3+3 (2 y^2-5 y+2)
x^2+y
(2 y^2-9 y+8)x \nonumber \\
&&~~~~~~~~~~~~-(y-2) y^2)
\end{eqnarray}
and
\begin{eqnarray}
C_3^{p\sigma}(2)=8 m_c^2, ~~~C_2^{p\sigma}(2)=-8 m_c^4 (4 x^2-8 y
x+4 y+3),\nonumber
\end{eqnarray}
\begin{eqnarray}
&&C_1^{p\sigma}(2)=16 m_c^6 (4 x^2+(3-14 y) x+7 y),\nonumber\\
&&C_0^{p\sigma}(2)=64 m_c^8 (x^4-2 x^3-3 (y-1) y x^2-2 y (y^2-3 y+1)
x+(y-2) y^2)
\end{eqnarray}
for hard-scattering amplitude
$T_{HJ}^{p\sigma}(x,y,\textbf{k}_1,\textbf{k}_2)$ with $J=0,2$.
\begin{eqnarray}
&&C_3^{\sigma p}(0)=5 m_c^2, ~~~C_2^{\sigma p}(0)=4 m_c^4 (4 y^2-6 y+x (4 y-2)-3),\nonumber\\
&&C_1^{\sigma p}(0)= -4 m_c^6 (3 x^2+2 (7 y-5) x+y (11 y-18)),\nonumber\\
&&C_0^{\sigma p}(0)=-16 m_c^8 ((2 y-1) x^3+(6 y^2-9 y+2) x^2+y (6
y^2-15 y+8) x\nonumber \\
&&~~~~~~~~~~~~+y^2 (2y^2-7 y+6))
\end{eqnarray}
and
\begin{eqnarray}
&&C_3^{\sigma p}(2)=8 m_c^2, ~~~C_2^{\sigma p}(2)=-8 m_c^4 (4 y^2+x (4-8 y)+3),\nonumber\\
&&C_1^{\sigma p}(2)=16 m_c^6 (x (7-14 y)+y (4 y+3)), \nonumber\\
&&C_0^{\sigma p}(2)=-64 m_c^8 ((2 y-1) x^3+(3 y^2-6 y+2) x^2+(2-3 y)
y x-(y-2) y^3)
\end{eqnarray}
for hard-scattering amplitude $T_{HJ}^{\sigma
p}(x,y,\textbf{k}_1,\textbf{k}_2)$ with $J=0,2$.
\begin{eqnarray}
&&C_6^{\sigma \sigma}(0)=7 m_c^2, ~~~C_5^{\sigma \sigma}(0)=-2 m_c^4 (7 x^2-7 x+11 y^2-11 y+39),\nonumber\\
&&C_4^{\sigma \sigma}(0)=2 m_c^6 ((26-52 y) x^3+(-104 y^2+182 y-101)
x^2-2 (26 y^3-91 y^2+173 \
y-85) x \nonumber\\
&&~~~~~~~~~~~~+26 y^3-69 y^2+138 y+40),\nonumber
\end{eqnarray}
\begin{eqnarray}
&&C_3^{\sigma \sigma}(0)=-16 m_c^8 (x^4+(8 y-6) x^3+2 (5 y^2-11 y-3)
x^2-2 (5 y^2+18 y-18) x-3 \
y^4 \nonumber\\
&&~~~~~~~~~~~~+6 y^3-6 y^2+28 y-6), \nonumber\\
&&C_2^{\sigma \sigma}(0)=32 m_c^{10} ((8 y-4) x^5+4 (8 y^2-13 y+5)
x^4+2
(24 y^3-68 y^2+62 y-19) \ x^3 \nonumber\\
&&~~~~~~~~~~~~+(32 y^4-136 y^3+200 y^2-130 y+35) x^2+2 (4 y^5-26
y^4+54 y^3-53 \
y^2+31 y \nonumber\\
&&~~~~~~~~~~~~-9) x+y (-4 y^4+12 y^3-14 y^2+19 y-18)), \nonumber\\
&&C_1^{\sigma \sigma}(0)=32 m_c^{12} (x^6+(1-8 y) x^5+(-45 y^2+65
y-9)
x^4-2 (40 y^3-105 y^2+50 \ y+6) x^3 \nonumber\\
&&~~~~~~~~~~~~+(-65 y^4+250 y^3-214 y^2-36 y+36) x^2+y (-24 y^4+125
y^3-164 \
y^2-20 y \nonumber\\
&&~~~~~~~~~~~~+72) x+y^2 (-3 y^4+21 y^3-41 y^2+4 y+36)),\nonumber\\
&&C_0^{\sigma \sigma}(0)=-96 m_c^{14} (x^2+2 (y-1) x+(y-2) y)^3 (-2
y+x (4 y-2)-1)
\end{eqnarray}
and
\begin{eqnarray}
&&C_6^{\sigma \sigma}(2)=4 m_c^2, ~~~C_5^{\sigma \sigma}(2)=4 m_c^4 (7 x^2-(12 y+1) x+11 y^2-5 y-6),\nonumber\\
&&C_4^{\sigma \sigma}(2)=4 m_c^6 (26 (2 y-1) x^3+(-208 y^2+130 y-61)
x^2+(52 y^3+130 y^2-32 \
y+25) x \nonumber\\
&&~~~~~~~~~~~~-26 y^3-93 y^2+57 y-4),\nonumber\\
&&C_3^{\sigma \sigma}(2)=32 m_c^8 (x^4+(21-46 y) x^3+(118 y^2-49
y+15) x^2-y (54 y^2+37 y+6) \
x-3 y^4 \nonumber\\
&&~~~~~~~~~~~~+33 y^3+15 y^2-8 y+6), \nonumber\\
&&C_2^{\sigma \sigma}(2)=-32 m_c^{10} (8 (2 y-1) x^5+(-32 y^2-8 y+7)
x^4+(-96 y^3+208 y^2-166 \ y+53) x^3 \nonumber\\
&&~~~~~~~~~~~~+(-32 y^4+208 y^3-26 y^2-29 y+10) x^2+(16 y^5-8
y^4-198 \
y^3+19 y^2 \nonumber\\
&&~~~~~~~~~~~~-56 y+30) x+y (-8 y^4-9 y^3+101 y^2-22 y+30)), \nonumber\\
&&C_1^{\sigma \sigma}(2)=-64 m_c^{12} (x^6+(13-32 y) x^5+(51 y^2+29
y-12) x^4+2 (80 y^3-171 \ y^2+64 y\nonumber\\
&&~~~~~~~~~~~~-12) x^3+(31 y^4-302 y^3+248 y^2+24 y-24) x^2+y (-48
y^4+89 \
y^3+64 y^2\nonumber\\
&&~~~~~~~~~~~~+40 y-48) x-y^2 (3 y^4-33 y^3+44 y^2+8 y+24)),\nonumber\\
&&C_0^{\sigma \sigma}(2)=192 m_c^{14} (x^2+2 (y-1) x+(y-2) y)^2 ((4
y-2) x^3+(-16 y^2+10 y-3) \
x^2+(4 y^3\nonumber\\
&&~~~~~~~~~~~~+10 y^2-1) x-y (2 y^2+3 y+1))
\end{eqnarray}
for hard-scattering amplitude $T_{HJ}^{\sigma
p}(x,y,\textbf{k}_1,\textbf{k}_2)$ with $J=0,2$.

\section{ the coefficients of the Fourier transform}

In this appendix we present the explicit expression of the
coefficients $A_{1i}^I(J),~A_{2i}^I(J)$ and $B_i^I(J)$ appearing in
the Eq.(16) for the Fourier transform of hard-scattering amplitude.
First, we show some useful formulas for Fourier transform,
\begin{eqnarray}
\int{d^2\textbf{k}\over(2\pi)^2}{e^{-i
\textbf{k}\cdot\textbf{b}}\over (s-\textbf{k}^2+i
\epsilon)^n}={(-)^n\over (n-1)!}\left \{
\begin{array}{l@{\quad~~\quad}r}
{i\over 4}{d^{n-1}\over ds^{n-1}}[H^{(1)}_0(\sqrt{s}b)] & \texttt{for}~~s>0 \\
{1\over 2\pi}{d^{n-1}\over ds^{n-1}}[K_0(\sqrt{-s}b)] &
\texttt{for}~~s<0
\end{array} \right.
\end{eqnarray}
with $n=1,2,\cdots$.

The coefficients $A_{1i}^I(J),~A_{2i}^I(J)$ and $B_i^I(J)$, coming
from the Fourier transform of hard-scattering amplitudes, are given
as,
\begin{eqnarray}
A_{10}^{\pi\pi}(J)={2(x+y)+(-2)^{J/2}(x-y)^2\over 8
(x+y-1)(x+y)^2},~~A_{20}^{\pi\pi}(J)={2(2-x-y)+(-2)^{J/2}(x-y)^2\over
8 (x+y-1)(2-x-y)^2},\nonumber
\end{eqnarray}
\begin{eqnarray}
B_0^{\pi\pi}(J)={(x+y)(2-x-y)+(-2)^{J/2}(x-y)^2 \over \pi
(x+y)^2(2-x-y)^2},\nonumber
\end{eqnarray}
\begin{eqnarray}
B_1^{\pi\pi}(J)={(-2)^{J/2}b(x-y)^2\over 4\sqrt{2}\pi
(x+y)(2-x-y)\sqrt{x+y-2xy}}
\end{eqnarray}
with $b=2 m_c b_1$ for $\mathcal
{T}_{HJ}^{\pi\pi}(x,y,\textbf{b}_1,\textbf{b}_2)$,
\begin{eqnarray}
A_{10}^{pp}(0)={x+y+3\over 4
(x+y-1)(x+y)},~~A_{20}^{pp}(0)={5-x-y\over 4
(x+y-1)(2-x-y)},\nonumber
\end{eqnarray}
\begin{eqnarray}
B_0^{pp}(0)={3 \over \pi (x+y)(2-x-y)},~~B_1^{pp}(0)=-{b\over
2\sqrt{2}\pi \sqrt{x+y-2xy}}
\end{eqnarray}
for $\mathcal {T}_{H0}^{pp}(x,y,\textbf{b}_1,\textbf{b}_2)$,
\begin{eqnarray}
A_{10}^{pp}(2)={4xy-x^2-y^2\over 2
(x+y-1)(x+y)^2},~~A_{20}^{pp}(2)={4(1-x)(1-y)-(1-x)^2-(1-y)^2\over 2
(x+y-1)(2-x-y)^2},\nonumber
\end{eqnarray}
\begin{eqnarray}
B_0^{pp}(2)={-6(x-y)^2 \over \pi
(x+y)^2(2-x-y)^2},~~B_1^{pp}(2)={-b((x-y)^2+x+y-2xy)\over
\sqrt{2}\pi(x+y)(2-x-y) \sqrt{x+y-2xy}}
\end{eqnarray}
for $\mathcal {T}_{H2}^{pp}(x,y,\textbf{b}_1,\textbf{b}_2)$,
\begin{eqnarray}
A_{10}^{p\sigma}(0)=-{4 x^2+(2 y-3) x-2 y^2+y\over 8(x+y-1)^2
(x+y)^3 },~~A_{20}^{p\sigma}(0)={4 x^2+(2 y-7) x-2 y^2+y+2\over
8(x+y-1)^2(2-x-y)^3},\nonumber
\end{eqnarray}
\begin{eqnarray}
A_{11}^{p\sigma}(0)={-b x\over
16(x+y-1)(x+y)\sqrt{xy}},~~A_{21}^{p\sigma}(0)={-b(1-x)\over 16
(x+y-1)(2-x-y)\sqrt{(1-x)(1-y)}}\nonumber
\end{eqnarray}
\begin{eqnarray}
B_0^{p\sigma}(0)&=&{1\over 32 \pi (x+y)^3(2-x-y)^3 (x+y-2xy)}(2 b^2
x^6+b^2 (10 y-11) x^5+5 b^2 (4 y^2-9 y\nonumber \\
&&+4) x^4+2 ((10 y^3-35 \ y^2+32 y-6) b^2-160 y+80) x^3+2 (5 b^2
y^4-25 b^2 y^3\nonumber\\
&&+4 (9 b^2-16) \ y^2+(304-14 b^2) y-96) x^2+y (2 b^2 y^4-15 b^2
y^3+32 (b^2+6) y^2\nonumber\\
&&-20 \ (b^2+8) y-128) x-y^2 (b^2 y^3-4 b^2 y^2+4 (b^2+24)
y-64))\nonumber
\end{eqnarray}
\begin{eqnarray}
B_1^{p\sigma}(0)={b (2 x^4+(26 y-17) x^3+(14 y^2-53 y+18) x^2+y (-10
y^2+y+16) x+y^2 (5 \ y-2))\over 8 \sqrt{2} \pi (2-x-y)^2 (x+y)^2
\sqrt{x+y-2xy} (2xy-x-y)}\nonumber
\end{eqnarray}
\begin{eqnarray}
B_2^{p\sigma}(0)={b^2 (2 x^2+(2 y-3) x-y)\over 32 \pi
(x+y)(2-x-y)(x+y-2xy)}
\end{eqnarray}
for $\mathcal {T}_{H0}^{p\sigma}(x,y,\textbf{b}_1,\textbf{b}_2)$,
\begin{eqnarray}
A_{10}^{p\sigma}(2)={4 x^2+(2 y-3) x-2 y^2+y\over 4(x+y-1)^2 (x+y)^3
},~~A_{20}^{p\sigma}(2)=-{4 x^2+(2 y-7) x-2 y^2+y+2\over
4(x+y-1)^2(2-x-y)^3},\nonumber
\end{eqnarray}
\begin{eqnarray}
A_{11}^{p\sigma}(2)={b x (x-2y)\over
8(x+y-1)(x+y)^2\sqrt{xy}},~~A_{21}^{p\sigma
}(2)={b(x-1)(x-2y+1)\over
8(x+y-1)(2-x-y)^2\sqrt{(1-x)(1-y)}}\nonumber
\end{eqnarray}
\begin{eqnarray}
B_0^{p\sigma}(2)&=&{1\over 8 \pi (x+y)^3(2-x-y)^3 (x+y-2xy)}(-b^2
x^6-2 b^2 (y-2) x^5+b^2 (2 y^2+3 y-4) x^4\nonumber \\
&&+4 (2 b^2 y^3-4 b^2 \ y^2+(b^2+40) y-20) x^3+(7 b^2 y^4-26 b^2
y^3+8 (3 b^2+8) y^2\nonumber \\
&&-4 \ (b^2+76) y+96) x^2+2 y (b^2 y^4-6 b^2 y^3+2 (5 b^2-24) y^2-4
(b^2-10) \ y+32) x\nonumber \\
&&-y^2 (b^2 y^3-4 b^2 y^2+4 (b^2-12) y+32))\nonumber
\end{eqnarray}
\begin{eqnarray}
B_1^{p\sigma}(2)={-b (x^4+2 (8 y-5) x^3+(-11 y^2-13 y+6) x^2-2 y
(y^2-7 y-1) x+(y-4) y^2)\over 2 \sqrt{2} \pi (2-x-y)^2 (x+y)^2
\sqrt{x+y-2xy} (2xy-x-y)}\nonumber
\end{eqnarray}
\begin{eqnarray}
B_2^{p\sigma}(2)={-b^2 (x^2-2xy+y)\over 8 \pi (x+y)(2-x-y)(x+y-2xy)}
\end{eqnarray}
for $\mathcal {T}_{H2}^{p\sigma}(x,y,\textbf{b}_1,\textbf{b}_2)$,
\begin{eqnarray}
A_{10}^{\sigma p}(0)={2 x^2-(2 y+1) x+(3-4 y) y\over 8(x+y-1)^2
(x+y)^3 },~~A_{20}^{\sigma p}(0)={-2 x^2+2 y x+x+4 y^2-7 y+2\over
8(x+y-1)^2(2-x-y)^3},\nonumber
\end{eqnarray}
\begin{eqnarray}
A_{11}^{\sigma p}(0)={-b y\over
16(x+y-1)(x+y)\sqrt{xy}},~~A_{21}^{\sigma p}(0)={-b(1-y)\over 16
(x+y-1)(2-x-y)\sqrt{(1-x)(1-y)}}\nonumber
\end{eqnarray}
\begin{eqnarray}
B_0^{\sigma p}(0)&=&{1\over 32 \pi (x+y)^3(2-x-y)^3 (x+y-2xy)}(b^2
(2 y-1) x^5+b^2 (10 y^2-15 y+4) x^4\nonumber\\
&&+2 ((10 y^3-25 y^2+16 y-2) \ b^2+96 y-48) x^3+2 (10 b^2 y^4-35 b^2
y^3+4 (9 b^2\nonumber\\
&&-16) y^2-10 (b^2+8) \ y+32) x^2+y (10 b^2 y^4-45 b^2 y^3+64
(b^2-5) y^2+(608\nonumber \\
&&-28 b^2) y-128) \ x+y^2 (2 b^2 y^4-11 b^2 y^3+20 b^2 y^2-4 (3
b^2-40) y-192))\nonumber
\end{eqnarray}
\begin{eqnarray}
B_1^{\sigma p}(0)={b ((5-10 y) x^3+(14 y^2+y-2) x^2+y (26 y^2-53
y+16) x+y^2 (2 y^2-17 \ y+18))\over 8 \sqrt{2} \pi (2-x-y)^2 (x+y)^2
\sqrt{x+y-2xy} (2xy-x-y)}\nonumber
\end{eqnarray}
\begin{eqnarray}
B_2^{\sigma p}(0)={b^2 (y (2 y-3)+x (2 y-1))\over 32 \pi
(x+y)(2-x-y)(x+y-2xy)}
\end{eqnarray}
for $\mathcal {T}_{H0}^{\sigma p}(x,y,\textbf{b}_1,\textbf{b}_2)$,
\begin{eqnarray}
A_{10}^{\sigma p}(2)={-2 x^2+2 y x+x+y (4 y-3)\over 4(x+y-1)^2
(x+y)^3 },~~A_{20}^{\sigma p}(2)=-{-2 x^2+2 y x+x+4 y^2-7 y+2\over
4(x+y-1)^2(2-x-y)^3},\nonumber
\end{eqnarray}
\begin{eqnarray}
A_{11}^{\sigma p}(2)={b y (y-2x)\over
8(x+y-1)(x+y)^2\sqrt{xy}},~~A_{21}^{\sigma
p}(2)={b(y-1)(y-2x+1)\over
8(x+y-1)(2-x-y)^2\sqrt{(1-x)(1-y)}}\nonumber
\end{eqnarray}
\begin{eqnarray}
B_0^{\sigma p}(2)&=&{1\over 8 \pi (x+y)^3(2-x-y)^3 (x+y-2xy)}(b^2 (2
y-1) x^5+b^2 (7 y^2-12 y+4) x^4\nonumber \\
&&+((8 y^3-26 y^2+20 y-4) b^2-96 \ y+48) x^3+2 (b^2 y^4-8 b^2 y^3+4
(3
b^2+8) y^2\nonumber \\
&&-4 (b^2-10) y-16) x^2+y \ (-2 b^2 y^4+3 b^2 y^3+4 (b^2+40) y^2-4
(b^2+76) y+64) x\nonumber \\
&&-y^2 (b^2 \ y^4-4 b^2 y^3+4 b^2 y^2+80 y-96))\nonumber
\end{eqnarray}
\begin{eqnarray}
B_1^{\sigma p}(2)={b ((2 y-1) x^3+(11 y^2-14 y+4) x^2+y (-16 y^2+13
y-2) x-y^2 (y^2-10 \ y+6))\over 2 \sqrt{2} \pi (2-x-y)^2 (x+y)^2
\sqrt{x+y-2xy} (2xy-x-y)}\nonumber
\end{eqnarray}
\begin{eqnarray}
B_2^{\sigma p}(2)={-b^2 (y^2-2xy+x)\over 8 \pi
(x+y)(2-x-y)(x+y-2xy)}
\end{eqnarray}
for $\mathcal {T}_{H2}^{\sigma p}(x,y,\textbf{b}_1,\textbf{b}_2)$,
\begin{eqnarray}
A_{10}^{\sigma\sigma}(0)&=&{1\over 128(x+-1)^3(x+y)^4}(b^2 x^6+((6
y-1) b^2-96 y+64) x^5+(b^2 (15 y^2-5 y-1)\nonumber \\
&&-8 (48 y^2-36 \ y+1)) x^4+(b^2 (20 y^3-10 y^2-4 y+1)-48 (12 y^3-10
y^2+4 y\nonumber \\
&&+1)) x^3+(3 \ (5 b^2-128) y^4+(352-10 b^2) y^3-2 (3 b^2+88) y^2+3
(b^2+48) y\nonumber \\
&&+24) \ x^2+y (6 (b^2-16) y^4+(96-5 b^2) y^3-4 (b^2-48) y^2+3
(b^2-16) y-80) \ x\nonumber \\
&&+y^2 (b^2 y^4-b^2 y^3-(b^2-184) y^2+(b^2-240) y+88))\nonumber
\end{eqnarray}
\begin{eqnarray}
A_{20}^{\sigma\sigma}(0)&=&{-1\over 128 (x+y-1)^3(2-x-y)^4}(8 (4 (3
y-1) x^5+3 (16 y^2-40 y+11) x^4+2 (36 y^3\nonumber \\
&&-174 y^2+240 y-67) \ x^3+(48 y^4-364 y^3+934 y^2-926 y+249) x^2+2
(6 y^5\nonumber \\
&&-72 y^4+280 \ y^3-499 y^2+397 y-100) x-12 y^5+73 y^4-206 y^3+305
y^2-216 y\nonumber \\
&&+52)-b^2 \ (x+y-3) (x+y-2)^3 (x+y-1)^2)\nonumber
\end{eqnarray}
\begin{eqnarray}
A_{11}^{\sigma\sigma}(0)&=& {b\over 64 (x+y-1)^2(x+y)^3\sqrt{xy}}
((1-8 y) x^4+(-24 y^2-28 y+6) x^3-(24 y^3+58 y^2\nonumber \\
&&-34 y+3) x^2+2 y \ (-4 y^3-14 y^2+17 y+1) x+y^2 (y^2+6
y-3))\nonumber
\end{eqnarray}
\begin{eqnarray}
A_{21}^{\sigma\sigma}(0)&=&{b\over 64
(x+y-1)^2(2-x-y)^3\sqrt{(1-x)(1-y)}}((7-8 y) x^4-2 (12 y^2-54 y+37)
x^3\nonumber \\
&&+(-24 y^3+202 y^2-430 y+231) x^2-2 \ (4 y^4-54 y^3+215 y^2-307
y+134) x\nonumber \\
&&+7 y^4-74 y^3+231 y^2-268 y+100)\nonumber
\end{eqnarray}
\begin{eqnarray}
A_{12}^{\sigma\sigma}(0)={-b^2 (x+y+1)\over 128 (x+y)(x+y-1)}
,~~A_{22}^{\sigma\sigma}(0)={b^2 (x+y-3)\over 128
(2-x-y)(x+y-1)}\nonumber
\end{eqnarray}
\begin{eqnarray}
B_0^{\sigma\sigma}(0)&=&{1\over 256\pi
(x+y)^4(2-x-y)^4(x+y-2xy)^2}(10 b^2 (2 y-1) x^9+(b^2 (336 y^2-426
y\nonumber \\
&&+121)-128 (1-2 y)^2) x^8+8 \ ((198 y^3-465 y^2+309 y-59) b^2+16
(-16 y^3\nonumber \\
&&+40 y^2-26 y+5)) x^7+4 \ (b^2 (892 y^4-3170 y^3+3571 y^2-1482
y+186)\nonumber \\
&&-32 (12 y^4-80 y^3+72 \ y^2-11 y-3)) x^6+4 ((1110 y^5-5451
y^4+9006 y^3\nonumber \\
&&-6006 y^2+1516 y-96) \ b^2+32 (32 y^5-44 y^4-110 y^3+45 y^2+34
y-12)) x^5\nonumber \\
&&+2 ((1560 y^6-10230 \ y^5+23235 y^4-22260 y^3+8620 y^2-960 y-24)
b^2\nonumber \\
&&+64 (68 y^6-284 y^5+246 \ y^4+203 y^3-115 y^2-28 y+12)) x^4+8 y
((142 y^6\nonumber \\
&&-1277 y^5+3947 \ y^4-5225 y^3+2868 y^2-464 y-24) b^2+16 (48
y^6-304 y^5\nonumber
\end{eqnarray}
\begin{eqnarray}
&&+730 y^4-537 \ y^3-196 y^2+168 y-16)) x^3+4 y^2 ((36 y^6-570
y^5+2603 y^4\nonumber \\
&&-4878 \ y^3+3782 y^2-864 y-72) b^2+32 (12 y^6-120 y^5+496 y^4-967
y^3\nonumber \\
&&+765 \ y^2-88 y-24)) x^2-2 y^3 ((6 y^6+45 y^5-612 y^4+1924
y^3-2264 y^2+768 \ y\nonumber \\
&&+96) b^2+64 (12 y^5-78 y^4+281 y^3-482 y^2+348 y-48)) x+y^4 (b^2
(6 \ y^3+33 y^2\nonumber \\
&&-76 y-12) (y-2)^2+128 (3 y^4-15 y^3+51 y^2-76 y+44)))\nonumber
\end{eqnarray}
\begin{eqnarray}
B_1^{\sigma\sigma}(0)&=&{-b\over 512\sqrt{2}\pi
(x+y)^3(2-x-y)^3(x+y-2xy)^{5/2}}(-6 b^2 (1-2 y)^2 x^8-(2 y-1)\nonumber \\
&& (3 b^2 (24 y^2-40 y+11)-16 (32 y^2-32 \ y+13)) x^7+(8 (512
y^4-1472 y^3+1224 y^2\nonumber \\
&&-446 y+61)-3 b^2 (120 y^4-408 \ y^3+428 y^2-161 y+18)) x^6+(16
(384 y^5\nonumber \\
&&-1728 y^4+3318 y^3-2541 \ y^2+863 y-103)-3 b^2 (160 y^5-760
y^4+1186 y^3\nonumber \\
&&-715 y^2+148 y-4)) \ x^5+(8 (512 y^6-3456 y^5+10480 y^4-16590
y^3+11011 y^2\nonumber \\
&&-2982 y+236)-3 \ b^2 (120 y^6-760 y^5+1636 y^4-1417 y^3+430 y^2-4
y-8)) x^4\nonumber \\
&&+(-16 (9 \ b^2-64) y^7+8 (153 b^2-1472) y^6+(43616-3558 b^2)
y^5+(4251 \ b^2\nonumber \\
&&-109040) y^4-24 (75 b^2-5836) y^3-8 (3 b^2+8428) y^2+96 (b^2+100)
\ y+192) x^3\nonumber \\
&&+y (-24 b^2 y^7+24 (13 b^2-64) y^6+(4672-1284 b^2) y^5+33 \ (65
b^2-336) y^4\nonumber \\
&&+(42904-1290 b^2) y^3-24 (b^2+2084) y^2+48 (3 \ b^2+268) y+576)
x^2\nonumber \\
&&+y^2 (24 b^2 y^6+(672-186 b^2) y^5+3 (161 b^2+816) \ y^4-4 (111
b^2+1796) y^3+12 (b^2\nonumber\\
&&-196) y^2+32 (3 b^2+140) y+576) x+y^3 \ (-6 b^2 y^5+(33 b^2-80)
y^4-6 (9 b^2\nonumber\\
&&+196) y^3+12 (b^2+204) y^2+24 \ (b^2-28) y+192))\nonumber
\end{eqnarray}
\begin{eqnarray}
B_2^{\sigma\sigma}(0)&=&{b^2\over 256 \pi
(x+y)^2(2-x-y)^2(x+y-2xy)^2}(10 (2 y-1) x^5+(256 y^2-306 y+81) x^4\nonumber \\
&&+4 (110 y^3-293 y^2+181 y-27) \ x^3+2 (96 y^4-522 y^3+619 y^2-162
y-6) x^2\nonumber \\
&&-2 y (6 y^4+81 y^3-282 \ y^2+146 y+12) x+y^2 (6 y^3+33 y^2-76
y-12))\nonumber
\end{eqnarray}
\begin{eqnarray}
B_3^{\sigma\sigma}(0)={b^3(-2 y+x (4 y-2)-1)\over 512\sqrt{2}\pi
(x+y-2xy)\sqrt{x+y-2xy}}
\end{eqnarray}
for $\mathcal{T}_{H0}^{\sigma
\sigma}(x,y,\textbf{b}_1,\textbf{b}_2)$ and
\begin{eqnarray}
A_{10}^{\sigma\sigma}(2)&=&{1\over 64 (x+y-1)^3(x+y)^4}(-b^2
x^6+(b^2+96 y-64) x^5+(b^2 (9 y^2-7 y+1)-8 (24 y^2\nonumber \\
&&+36 y-19)) \ x^4+((16 y^3-26 y^2+10 y-1) b^2+24 (-24 y^3+28 y^2+14
y-3)) x^3\nonumber\\
&&+(3 (3 \ b^2-64) y^4+(800-26 b^2) y^3+2 (9 b^2-200) y^2-3 (b^2+8)
y-24) \ x^2\nonumber
\end{eqnarray}
\begin{eqnarray}
&&+(96 y^5-(7 b^2+96) y^4+2 (5 b^2-24) y^3-3 (b^2-56) y^2-64 y+24) \
x+y (-b^2 y^5\nonumber \\
&&+b^2 y^4+(b^2-40) y^3-(b^2-120) y^2-88 y+24))\nonumber
\end{eqnarray}
\begin{eqnarray}
A_{20}^{\sigma\sigma}(2)&=&{-1\over 64()(x+y-1)^3(2-x-y)^4}(b^2
x^6+(-5 b^2-96 y+32) x^5+((-9 y^2+11 y\nonumber \\
&&+7) b^2+24 (8 y^2-8 y+7)) \ x^4+((-16 y^3+58 y^2-50 y+1) b^2+8 (72
y^3-228 y^2\nonumber \\
&&+222 y-79)) \ x^3+((-9 y^4+58 y^3-114 y^2+75 y-8) b^2+8 (24
y^4-212 y^3\nonumber \\
&&+434 y^2-301 \ y+51)) x^2+((11 y^4-50 y^3+75 y^2-40 y+4) b^2+8
(-12 y^5\nonumber \\
&&+142 y^3-229 \ y^2+64 y+41)) x+b^2 y^6+(96-5 b^2) y^5+(7 b^2-152)
y^4+(b^2\nonumber \\
&&-56) y^3-8 \ (b^2+5) y^2+4 (b^2+114) y-320)\nonumber
\end{eqnarray}
\begin{eqnarray}
A_{11}^{\sigma\sigma}(2)&=&{b\over 32(x+y-1)^2(x+y)^3\sqrt{xy}}((8
y-1) x^4-2 (12 y^2+7 y-3) x^3+(-24 y^3+22 y^2\nonumber \\
&&+20 y-3) x^2+2 y (4 \ y^3-7 y^2+10 y-1) x-y^2 (y^2-6
y+3))\nonumber
\end{eqnarray}
\begin{eqnarray}
A_{21}^{\sigma\sigma}(2)&=&{b\over
32(x+y-1)^2(2-x-y)^3\sqrt{(1-x)(1-y)}}((8 y-7) x^4+(-24 y^2+30 y-4)
x^3\nonumber \\
&&+(-24 y^3+122 y^2-146 y+39) x^2+2 (4 \ y^4+15 y^3-73 y^2+71 y-13)
x-7 y^4\nonumber \\
&&-4 y^3+39 y^2-26 y-4)\nonumber
\end{eqnarray}
\begin{eqnarray}
A_{12}^{\sigma\sigma}(2)={b^2 (x^2-4 y x+x+y^2+y)\over
64(x+y-1)(x+y)^2} ,~~A_{22}^{\sigma\sigma}(2)={b^2 (x^2-4 y
x+x+y^2+y)\over 64 (x+y-1)(2-x-y)^2},\nonumber
\end{eqnarray}
\begin{eqnarray}
B_0^{\sigma\sigma}(2)&=&{1\over
128\pi(x+y)^4(2-x-y)^4(x+y-2xy)^2}(10 b^2 (1-2 y) x^9+((-72 y^2+162
y\nonumber \\
&&-55) b^2+128 (1-2 y)^2) x^8+((192 \ y^3-210 y+49) b^2+128 (160
y^3-256 y^2\nonumber \\
&&+134 y-23)) x^7+((1160 y^4-2992 \ y^3+2240 y^2-765 y+150) b^2+128
(12 y^4\nonumber \\
&&-584 y^3+732 y^2-293 y+36)) \ x^6+(b^2 (1992 y^5-8460 y^4+11538
y^3-6591 y^2\nonumber \\
&&+1868 y-228)-128 (320 \ y^5-764 y^4-290 y^3+747 y^2-284 y+30))
x^5\nonumber \\
&&+(b^2 (1608 y^6-9804 \ y^5+19998 y^4-16821 y^3+6250 y^2-996
y+24)\nonumber \\
&&-128 y (68 y^5-1004 \ y^4+2142 y^3-799 y^2-124 y+62)) x^4+(128 (5
b^2+96) y^7\nonumber \\
&&-16 (341 \ b^2+1600) y^6+2 (7993 b^2-35200) y^5+(196992-19541 b^2)
y^4+(9960 \ b^2\nonumber \\
&&-80896) y^3-8 (247 b^2-1728) y^2+32 (3 b^2-224) y+1536) x^3+y (24
\ (5 b^2-64) y^7\nonumber
\end{eqnarray}
\begin{eqnarray}
&&-96 (15 b^2+128) y^6+32 (191 b^2+656) y^5+(33920-11103 \ b^2)
y^4+(8362 b^2\nonumber \\
&&-96768) y^3-8 (279 b^2-5824) y^2+144 (b^2-128) \ y+4608) x^2+y^2
(12 b^2 y^7\nonumber \\
&&-6 (29 b^2-256) y^6+6 (173 b^2+640) \ y^5-(2845 b^2+128) y^4+4
(851 b^2-5248) y^3\nonumber \\
&&+(33024-1380 b^2) y^2+96 \ (b^2-160) y+4608) x-y^3 (6 b^2
y^6+(384-57 b^2) y^5\nonumber \\
&&+(239 b^2+384) \ y^4+(1536-470 b^2) y^3+4 (89 b^2-1088) y^2-8 (3
b^2-512) y\nonumber \\
&&-1536))\nonumber
\end{eqnarray}
\begin{eqnarray}
B_1^{\sigma\sigma}(2)&=&{-b\over
256\sqrt{2}\pi(x+y)^3(2-x-y)^3(x+y-2xy)^{5/2}}(6 b^2 (1-2 y)^2
x^8-(2 y-1) (3 (16 y\nonumber \\
&&-5) b^2+16 (32 y^2-32 y+13)) \ x^7-(3 b^2 (72 y^4-144 y^3+64
y^2-13 y+3)\nonumber \\
&&-8 (256 y^4-64 y^3-384 \ y^2+374 y-91)) x^6+(8 (768 y^5-2688
y^4+1716 y^3\nonumber \\
&&+786 y^2-940 y+179)-3 \ b^2 (128 y^5-536 y^4+650 y^3-311 y^2+74
y-8)) x^5\nonumber \\
&&+(8 (256 y^6-2688 \ y^5+7616 y^4-5922 y^3+1475 y^2-297 y+124)-3
b^2 (72 y^6\nonumber \\
&&-536 y^5+1124 \ y^4-833 y^3+221 y^2-8 y-4)) x^4-(1024 y^7+(512-432
b^2) y^6\nonumber \\
&&+50 (39 \ b^2-464) y^5+(71056-2499 b^2) y^4+300 (3 b^2-176) y^3+16
(3 b^2+1165) \ y^2\nonumber \\
&&-48 (b^2+124) y+864) x^3+y (24 b^2 y^7-96 (b^2-16) y^6-64 (3 \
b^2-32) y^5+3 (311 b^2\nonumber \\
&&-7760) y^4+(56984-663 b^2) y^3-48 (b^2+751) \ y^2+24 (3 b^2+520)
y-2592) x^2\nonumber \\
&&+y^2 (-24 b^2 y^6+(78 b^2-672) y^5+(39 \ b^2-3024) y^4+(13472-222
b^2) y^3\nonumber\\
&&+24 (b^2-995) y^2+16 (3 b^2+692) \ y-2592) x+y^3 (6 b^2 y^5+(80-15
b^2) y^4-9 (b^2\nonumber \\
&&-104) y^3+24 (b^2-111) \ y^2+12 (b^2+296) y-864))\nonumber
\end{eqnarray}
\begin{eqnarray}
B_2^{\sigma\sigma}(2)&=&{b^2\over
128\pi(x+y)^2(2-x-y)^2(x+y-2xy)^2}((10-20 y) x^5+(8 y^2+42 y-15)
x^4\nonumber \\
&&+(280 y^3-436 y^2+218 y-51) x^3+(72 \ y^4-564 y^3+466 y^2-117 y+6)
x^2\nonumber \\
&&+y (12 y^4-102 y^3+378 y^2-149 y+12) \ x+y^2 (-6 y^3+33 y^2-83
y+6)) \nonumber
\end{eqnarray}
\begin{eqnarray}
B_3^{\sigma\sigma}(2)&=&{b^3 \over
256\sqrt{2}\pi(x+y)(2-x-y)(x+y-2xy)\sqrt{x+y-2xy}}((4 y-2) x^3+(-16
y^2\nonumber \\
&&+10 y-3) x^2+(4 y^3+10 y^2-1) x-y (2 y^2+3 \ y+1))
\end{eqnarray}
for $\mathcal {T}_{H2}^{\sigma
\sigma}(x,y,\textbf{b}_1,\textbf{b}_2)$
\section{the function $s(x,b,Q)$ in the sudakov factor}

In this appendix we present the explicit expression of the exponent
$s(x, b,Q)$ appearing in the Sudakov factor. Defining the variables,
\begin{eqnarray}
\hat{q}\equiv \ln{x Q\over \sqrt{2}\Lambda_{QCD}},~~~~~\hat{b}\equiv
\ln{1\over b \Lambda_{QCD}},
\end{eqnarray}
the exponent $s(x, b,Q)$ is presented up to next-to-leading-log
approximation \cite{PRD52}
\begin{eqnarray}
&&s(x,b,Q)={A^{(1)}\over 2 \beta_1}\left[
\hat{q}\ln\left({\hat{q}\over
\hat{b}}\right)-\hat{q}+\hat{b}\right]+{A^{(2)}\over 4
\beta_1^2}\left( {\hat{q}\over \hat{b}}-1\right)-\left[
{A^{(2)}\over 4 \beta_1^2}-{A^{(1)}\over 4 \beta_1}\ln\left ({e^{2
\gamma_E-1}\over 2}\right ) \right]
\ln\left({\hat{q}\over \hat{b}}\right) \nonumber \\
&& +{A^{(1)} \beta_2 \over 4 \beta_1^3}
\hat{q}\left[{\ln(2\hat{q})+1\over \hat{q}}-{\ln(2\hat{b})+1\over
\hat{b}}\right]+{A^{(1)}\beta_2\over 8 \beta_1^3}[\ln^2(2
\hat{q})-\ln^2(2
\hat{b})]\nonumber \\
&&+{A^{(1)}\beta_2\over 8 \beta_1^3}\ln\left({e^{2 \gamma_E-1}\over
2}\right)\left[{\ln(2\hat{q})+1\over
\hat{q}}-{\ln(2\hat{b})+1\over\hat{b}}\right]-{A^{(2)}\beta_2\over
16 \beta_1^4}\left[{2\ln(2\hat{q})+3\over
\hat{q}}-{2\ln(2\hat{b})+3\over\hat{b}}\right]\nonumber \\
&&-{A^{(2)}\beta_2\over 16\beta_1^4}{\hat{q}-\hat{b}\over
\hat{b}^2}\left[2\ln(2\hat{b})+1\right]+{A^{(2)}\beta_2^2\over
432\beta_1^6}{\hat{q}-\hat{b}\over \hat{b}^3}\left[9
\ln^2(2\hat{b})+6\ln(2\hat{b})+2\right]\nonumber \\
&&+{A^{(2)}\beta_2^2\over 1728\beta_1^6}\left[{18
\ln^2(2\hat{q})+30\ln(2\hat{q})+19\over\hat{q}^2}-{18\ln^2(2\hat{b})+30\ln(2\hat{b})+19\over\hat{b}^2}\right]
\end{eqnarray}
where the coefficients $\beta_i$ and $A^{(i)}$ are
\begin{eqnarray}
&&\beta_1={33-2 n_f\over 12},~~~~\beta_2={153-19 n_f\over
24},\nonumber \\
&&A^{(1)}={4\over 3},~~~~A^{(2)}={67\over 9}-{\pi^2\over 3}-{10\over
27}n_f+{8\over 3}\beta_1 \ln{e^{\gamma_E}\over 2},
\end{eqnarray}
with $\gamma_E$ the Euler constant.

The exponent $s(x, b,Q)$ is obtained under the condition that
$xQ/\sqrt{2}> 1/b$, i.e. the longitudinal momentum should be larger
than the transverse momentum. So $s(x, b,Q)$ is defined for
$\hat{q}\geq \hat{b}$, and set to zero for $\hat{q}<\hat{b}$. As a
similar treatment, the complete Sudakov factor $e^{-S}$ is set to
unity, if $e^{-S} > 1$, in the numerical analysis. This corresponds
to a truncation at large $k_T$, which spoils the on-shell
requirement for the light valence quarks. The quark lines with large
$k_T$ should be absorbed into the hard scattering amplitude, instead
of the wave functions.

\begin{acknowledgments}
This work was supported partially by the Natural Science Foundation
of China, Grant Number: 10747154, 10805009 and 10575083.
\end{acknowledgments}


\begin{thebibliography}{99}

\bibitem{HSA} S.J. Brodsky and G.P. Lepage, Phys. Rev. $\bf{D 22}$ (1980) 2157.

\bibitem{QustEP} N. Isgur and C. H. Llewellyn Smith, Nucl. Phys. $\bf{B 317}$ (1989) 526.

\bibitem{LiSterman} H. N. Li and G. Sterman, Nucl. Phys. $\bf{B 381}$ (1992) 129.

\bibitem{octet} J. Bolz, P. Koll and G.A. Schuler, Phys. Lett. $\bf{B 392}$ (1997) 198;
Eur. Phys. J. $\bf{C 2}$ (1998) 705.

\bibitem{tranF} S. J. Brodsky and G. P. Lepage, Phys. Rev. $\bf{D 24}$ (1981)
1808; A. Duncan and A. H. Mueller, Phys. Rev. $\bf{D 21}$(1980)1636;
P. Kroll, M. Raulfs, Phys. Lett. $\bf{B 387}$ (1996) 848.

\bibitem{alphastranF} Tsung-Wen Yeh, Phys. Rev. $\bf{D 66}$ (2002) 014002;
B. Melie, D. Muler, K. Passek-Kumerickib, Phys. Rev. $\bf{D 68}$
(2003) 014013.

\bibitem{hightw} S. S. Agaev, Phys. Rev. $\bf{D 69}$ (2004) 094010; Phys. Rev. $\bf{D
72}$ (2005) 114010, Erratum-ibid. $\bf{D 73}$ (2006) 059902.

\bibitem{pionF} F.G. Cao, Y.B. Dai, and C.S. Huang, Eur. Phys. J. $\bf{C 11}$ (1999)
501; Tao Huang, Xing-Gang Wu, Phys. Rev. $\bf{D 70}$ (2004) 093013.

\bibitem{KaonF} Xing-Gang Wu, Tao Huang, JHEP $\bf{0804}$ (2008) 043.

\bibitem{HLF} V. M. Belyaev, A. Khodjamirian, R. Ruckl, Z. Phys.
$\bf{C60}$ (1993) 349; Zheng-Tao Wei, Mao-Zhi Yang, Nucl. Phys.
$\bf{B 642}$ (2002) 263.

\bibitem{BPP} Han-Wen Huang, Cai-Dian Lu, Toshiyuki Morii, Yue-Long Shen, GeLiang
Song, Phys. Rev. $\bf{D73}$ (2006) 014011; Wei Wang, Yu-Ming Wang,
De-Shan Yang, Cai-Dian Lu, Phys. Rev. $\bf{D 78}$ (2008) 034011.

\bibitem{BHL} S. J. Brodsky, T. Huang, and G. P. Lepage, in Particles and
Fields-2, Proceedings of the Banff Summer Institute, Banff, Alberta,
1981, edited by A. Z. Capri and A. N. Kamal (Plenum, New York,1983),
p. 143; Tao Huang, Bo-Qiang Ma, and Qi-Xing Shen, Phys. Rev. $\bf{D
49}$(1994)1490.

\bibitem{DA1} V.M. Braun and I.B. Filyanov, Z. Phys. $\bf{C 44}$ (1989) 157; $\bf{C 48}$
(1990) 239.

\bibitem{DA2} P. Ball, JHEP $\bf{9901}$ (1999) 010; P. Ball, V. M.
Braun, A. Lenz, JHEP $\bf{0605}$ (2006) 004.

\bibitem{DA3} Tao Huang, Ming-Zhen Zhou, Xing-Hua Wu, Phys. Rev. $\bf{D
 70}$ (2004) 014013; Eur. Phys. J. $\bf{C 42}$ (2005) 271.

\bibitem{BSform} J.H.Kuhn,J.Kaplan,E.G.O.Safiani, Nucl. Phys. $\bf{B 157}$ (1979) 125.

\bibitem{Li} Makiko Nagashima, Hsiang-nan Li, Eur. Phys. J. $\bf{C 40}$ (2005) 395.

\bibitem{Beneke} M. Beneke, T. Feldmann, Nucl. Phys. $\bf{B 592}$ (2001) 3.

\bibitem{DM} A. Duncan and A. H. Mueller, Phys. Lett. $\bf{B 93}$ (1980) 119.

\bibitem{DAmoment1} V. L. Chernyak and A. R. Zhitnitsky, Nucl. Phys. $\bf{B 201}$ (1982) 492.

\bibitem{DAmoment2} N. G. Stefanis, W. Schroers, and H.-C. Kim, Phys. Lett. $\bf{B 449}$ (1999) 299.

\bibitem{DAmoment4} A. Khodjamirian, T. Mannel, and M. Melcher, Phys. Rev. $\bf{D 70}$ (2004) 094002.

\bibitem{DAmoment5} V. M. Braun and A. Lenz, Phys. Rev. $\bf{D 70}$ (2004) 074020.

\bibitem{DAmoment6} P. Ball and R. Zwicky, Phys. Lett. $\bf{B 633}$ (2006) 289.

\bibitem{DAmoment7} P. Ball and R. Zwicky, JHEP $\bf{02}$ (2006) 034.

\bibitem{piong1} P. Ball and R. Zwicky, Phys. Lett. $\bf{B 625}$ (2005)
225.

\bibitem{piong2} A. Schmedding and O. I. Yakovlev, Phys. Rev. $\bf{D 62}$ (2000) 116002.

\bibitem{piong3} A. P. Bakulev, S. V. Mikhailov, and N. G. Stefanis, Phys. Rev.
$\bf{D 73}$ (2006) 056002.

\bibitem{piong4} S. S. Agaev, Phys. Rev. $\bf{D 72}$ (2005) 114010.

\bibitem{pmod1} C. Quigg and J. L. Rosner, Phys. Rep. $\bf{56}$(1979) 167.

\bibitem{pmod2} W. Buchmuller and S. H. Tye, Phys. Rev.$\bf{D 24}$(1981)132.

\bibitem{fitcp} M. L. Mangano and A. Petrelli, Phys. Lett. $\bf{B
352}$ (1995) 445.

\bibitem{tpQCD1} R. Barbieri, M. Caffo, R. Gatto and E. Remiddi, Nucl. Phys. $\bf{B 192}$ (1981) 61.

\bibitem{tpQCD2} W. Kwong, P. B. Mackenzie, R. Rosenfeld and J. L.
Rosner, Phys. Rev. $ \bf{D 37}$ (1987) 3210.

\bibitem{tpQCD3} Han-Wen Huang and Kuang-Ta Chao, Phys. Rev. $\bf{D 54}$ (1996) 6850; Erratum-ibid. $\bf{D 56}$ (1997) 1821.

\bibitem{Cleo08} K.M. Ecklund et al.(CLEO Collaboration), Phys. Rev. $\bf{D 78}$ (2008) 091501.

\bibitem{pdg2008} C. Amsler et al. (Particle Data Group), Phys.
Lett. $\bf{ B 667}$ (2008)1.

\bibitem{Ap1} Cai-Dian Lu, Kazumasa Ukai, Mao-Zhi Yang, Phys. Rev. $\bf{D 63}$ (2001) 074009.

\bibitem{Ap2} Satoshi Mishima and A. I. Sanda, Prog. Theor. Phys. $\bf{110}$ (2003) 549.

\bibitem{Chtp} A. Pich, Talk given at Les Houches Summer School in Theoretical Physics, Session
68: Probing the Standard Model of Particle Interactions, Les
Houches, France, 28 Jul ¨C 5 Sep 1997, Preprint hep-ph/9806303.

\bibitem{BES98} J.Z. Bai et al. (BES Collaboration), Phys. Rev. Lett. $\bf{81}$ (1998) 3091.

\bibitem{Belle05} H. Nakazawa et al. (Belle Collaboration), Phys.
Lett. $\bf{B 615}$ (2005) 39.


\bibitem{PRD52} Hsiang-nan Li,  Phys. Rev. $\bf{D 52}$ (1995) 3958.

\end{thebibliography}
\end{document}